\documentclass[aps,prb,twocolumn,showpacs,floatfix]{revtex4}

\usepackage{graphicx}
\usepackage{dcolumn}
\usepackage[english]{babel}

\begin{document}

\title{{\textit{Molecular-Spintronics}}: the art of driving spin through molecules}
\author{S. Sanvito}
\email{sanvitos@tcd.ie}
\author{A. R. Rocha}
\affiliation{School of Physics, Trinity College, Dublin 2, IRELAND}
\date{\today}

\begin{abstract}
{\it Spintronics} is the ability of injecting, manipulating and detecting
electron spins into solid state systems. Molecular-electronics investigates
the possibility of making electronic devices using organic molecules. Traditionally
these two burgeoning areas have lived separate lives, but recently 
a growing number of experiments have indicated a possible pathway towards their
integration. This is the playground for {\it molecular-spintronics}, where
spin-polarized currents are carried through molecules, and in turn they can
affect the state of the molecule. We review the most recent advances
in molecular-spintronics. In particular we discuss how a fully quantitative
theory for spin-transport in nanostructures can offer fundamental insights
into the main factors affecting spin-transport at the molecular level, and
can help in designing novel concept devices.
\end{abstract}


\maketitle

\tableofcontents

\section{Introduction}\label{intro}
Very few scientific discoveries have moved from an academic laboratory to industrial
mass production as quickly as the giant magnetoresistance effect (GMR) \cite{GMR1,GMR2}, 
now exploited in any read-head for standard hard drives. 
GMR is the change of the electrical resistance
of a magnetic device when an external magnetic field is applied and it is essentially associated
to a change in the magnetic state of the device itself. The revolutionary scientific
message revealed by the GMR effect is that the electron spin, as well as the electronic
charge, can be used in electronic applications. This somehow has set a new paradigm.

More recently, the electron spin has made
its appearance in semiconductor physics. This new field, usually called spin-electronics
or spintronics \cite{spintronics_wolf,spintronics_prinz1,spintronics_prinz2} has the potential
of bringing memory and logic functionalities on the same chip. The electron spin is the ultimate 
logic bit. In semiconductors spin preserves coherence over extremely long times \cite{Aws1}
and distances \cite{Aws2}, thus it offers the tantalizing prospect of being used for
quantum logic \cite{spin_computer}. Moreover all electronic ways of manipulating the
spin direction have been proposed \cite{dasdatta}. These are based on the spin-orbit 
interaction \cite{rashba}, which interestingly plays a ubiquitous r\^ole in semiconductor spintronics.

One the one hand spin-orbit allows us to manipulate the electron spins by electric only means. It
is an intrinsic property of the electronic structure, and therefore it can be engineered by appropriate 
heterojunction fabrication and manipulated by stress or with an external electric field.
Importantly it can be controlled, at least in principle, at an extremely local level. The spin Hall 
\cite{spinhall} effect, a solid state version of the Stern-Gerlach experiment utilizing spin-orbit 
interaction instead of a magnetic field gradient, is a good example of all-electrical
spin manipulation. On the other hand spin-orbit is the main source of spin-dephasing through 
the DÕyakonov-PerelÕ mechanism \cite{spindep}, i.e. in semiconductors it is the 
main interaction responsible for reducing spin coherence.

Ultimately spin looks like an attractive degree of freedom to be used in logic because 
the energy scale relevant for its typical dynamics is order of magnitudes smaller than that involved in
manipulating the electron charge in standard transistors. This can translate in devices
exhibiting ultra-low power consumption and high speed. Moreover the sole existence of 
magnetic materials with high Curie temperature suggests the possibility of powerless 
non-volatility. 

At the same time and almost in parallel there has been a growing interest in making
electronic devices using organic molecules. This field, which takes the suggestive name
of molecular-electronics \cite{moltronics}, aims at replacing standard semiconductors with organic
materials. These have the advantages to be manufactured with low-temperature low-cost
chemical methods, instead of expensive high-temperature solid-state growth (e.g. molecular
beam epitaxy) and patterning (lithography) techniques. In addition the endless possibilities of
chemical synthesis and end-groups engineering give good expectation for new concept
devices. Negative differential resistance \cite{mol_ndr} and rectification 
\cite{mol_rec} have been already proved at the molecular level and prototypes of molecular 
transistors \cite{mol_trans}, memories \cite{memory} and logic gates \cite{gates1,gates2}, 
have all been demonstrated.

It is only until recently that spin has entered the realm of molecular electronics. The
driving idea behind the first pioneering experiment of Tsukagoshi and coworkers \cite{bruce},
who injected spin polarized electrons into carbon nanotubes, is that spin-orbit interaction is very 
weak in carbon-based materials. This fact, in addition to the rather weak hyperfine interaction, 
suggests extremely long spin relaxation times, and therefore the possibility of coherent
spin propagation over large distances. A rather conservative estimate of the spin diffusion length
from the Tsukagoshi's experiment indicated 130~nm as a lower bound of the spin-diffusion in
carbon nanotubes. These findings have stimulated a growing activity in the area and
several experiments dealing with molecular tunneling junctions \cite{Ralph}, spin-transport 
through polymers \cite{Shi,Dediu} and optical pump/probe experiments through 
molecular bridges \cite{Aws3} have recently appeared.

The molecular world has all the ingredients that spin-electronics needs. The conductivity
of polymers can be changed by more than ten order of magnitudes \cite{polymers} and
elementary molecules can be designed with the desired electronic structure. Molecules
can be anchored to metals in numerous ways and the bonding angle can be further engineered by
the coverage density. The spin-relaxation times can be extremely long and furthermore
both paramagnetic and ferromagnetic molecules \cite{roberta} are available. 

Theory and modeling is a powerful engine for the development of this new area. At present
accurate quantitative algorithms for evaluating the $I$-$V$ characteristics of molecular devices
are available \cite{Smeagol1,Smeagol2,NEGF1,NEGF2,NEGF3,NEGF4,NEGF5}, and they are
revolutionarizing the world of nanoscale device simulators, as density functional theory (DFT) \cite{DFT}
did for electronic structure methods in the sixties. Some of these algorithms are spin polarized 
\cite{Smeagol1,Smeagol2,NEGF1,NEGF4}, and therefore readily applicable to spin-transport 
phenomena. Certainly such calculations are not easy. For instance the degree of accuracy needed
for the description of the underling electronic structure may go beyond what is standard
in solid state physics \cite{SmeagolSIC}. Here in fact one needs to describe the
metallic state of the current/voltage electrodes, the molecular state of the actual device and 
magnetism on the same footing. In addition since a transport problem is essentially a non-equilibrium
problem, variational principles are not valid. One cannot depend on the free energy for atomic 
relaxation and the full dynamics must be considered \cite{Maria}.
Finally detailed information about the elementary 
excitations (phonons, spin waves etc.) and the exact atomic positions are essential.

The aim of this review is to offer a complete overview of the fascinating field of
molecular-spintronics. In particular we will demonstrate that a quantitative theory of 
quantum transport can offer important insights and can be an invaluable tool for
understanding complicated experiments and for novel device designing.
The paper is organized as follows. In the first two sections we will introduce the two
fields of spin- and molecular-electronics. Then we will introduce the main computational tools, 
and we will discuss the latest progresses with molecular spin-valves. Finally we
will overview the most recent and controversial findings, namely transport in
molecular magnets, d$^0$ ferromagnetism and contact induced ferromagnetism.

\section{Spin-Electronics}
\subsection{Transition metals and spin valves}
Magnetic transition metals and their permalloys occupy an important place in the field of 
spin-electronics. This is essentially due to their generally high Curie temperature (for a commercially 
useful magnetic materials it must exceed $\sim$500~$^o$K
\cite{PhysToday}), and the possibility of engineering the various magnetic properties by
alloying. The ferromagnetism in 3$d$ transition metals can be understood by simply looking at 
their electronic structure. 

The nominal atomic configurations of Ni, Co and Fe are respectively $4s^23d^8$, $4s^23d^7$ 
and $4s^23d^6$. Therefore in forming a solid one expects the Fermi level ($E_\mathrm{F}$) to be in a 
region of density of states (DOS) with dominant $d$ character. Since the $d$ shells are
rather localized the DOS is extremely large around $E_\mathrm{F}$ and the 
material becomes Stoner unstable thus developing a ferromagnetic ground state. 
The band energies $\epsilon_{\vec{k}\sigma}$ for the two different spin orientations 
($\sigma=\uparrow,\downarrow$) are shifted with respect to each 
other by a constant  $\Delta=\epsilon_{\vec{k}\downarrow}-\epsilon_{\vec{k}\uparrow}$, with
$\Delta$ approximately 1.4~eV in Fe, 1.3~eV in Co and 1.0~eV in Ni. More sophisticated DFT 
calculations show that such picture is a good approximation of the real electronic structure of 
Ni, Co and Fe. 

\begin{figure}[htb]
\begin{center}
\includegraphics[width=6.2cm,clip=true]{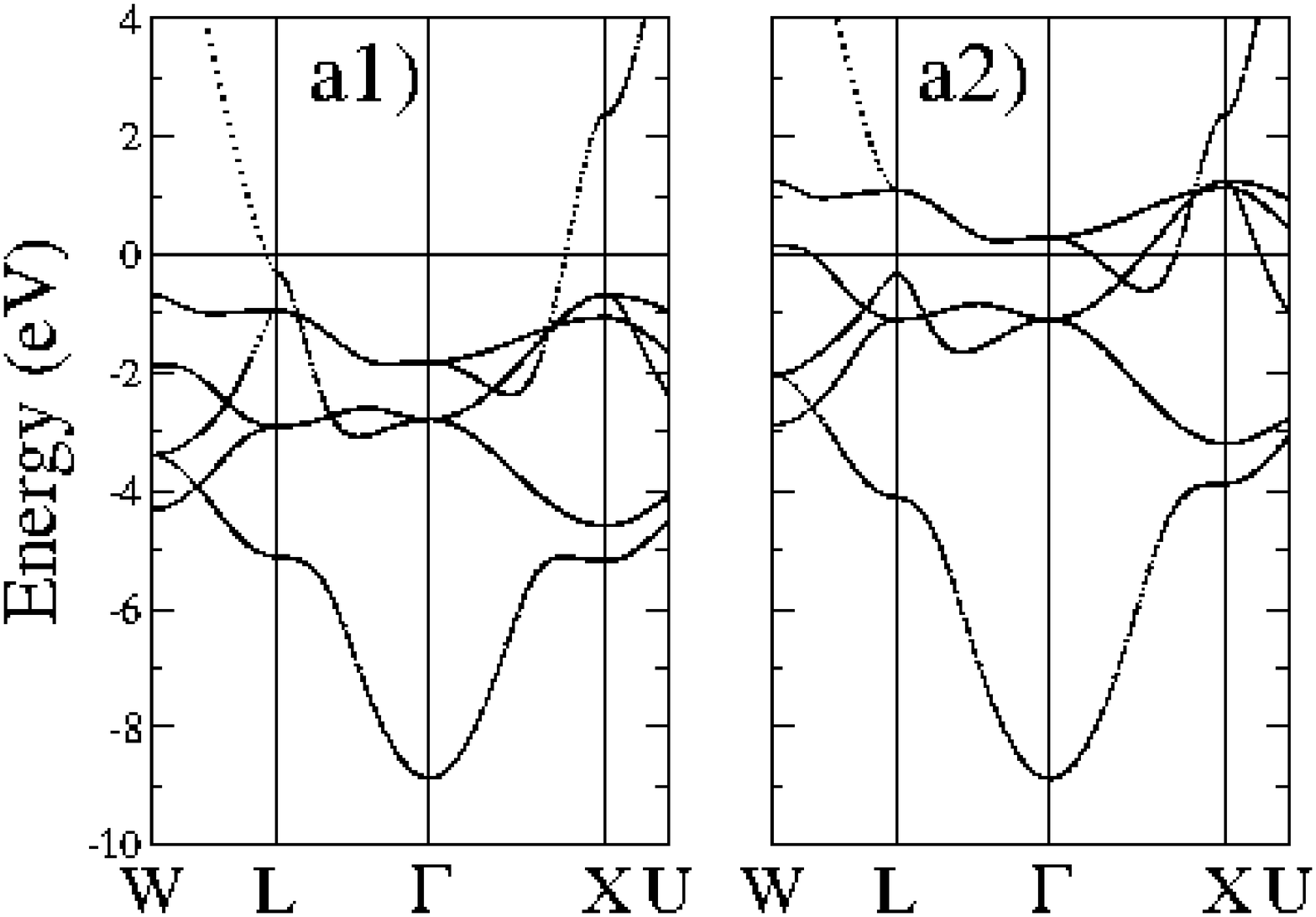}
\vspace{0.5cm}
\includegraphics[width=5.5cm,clip=true]{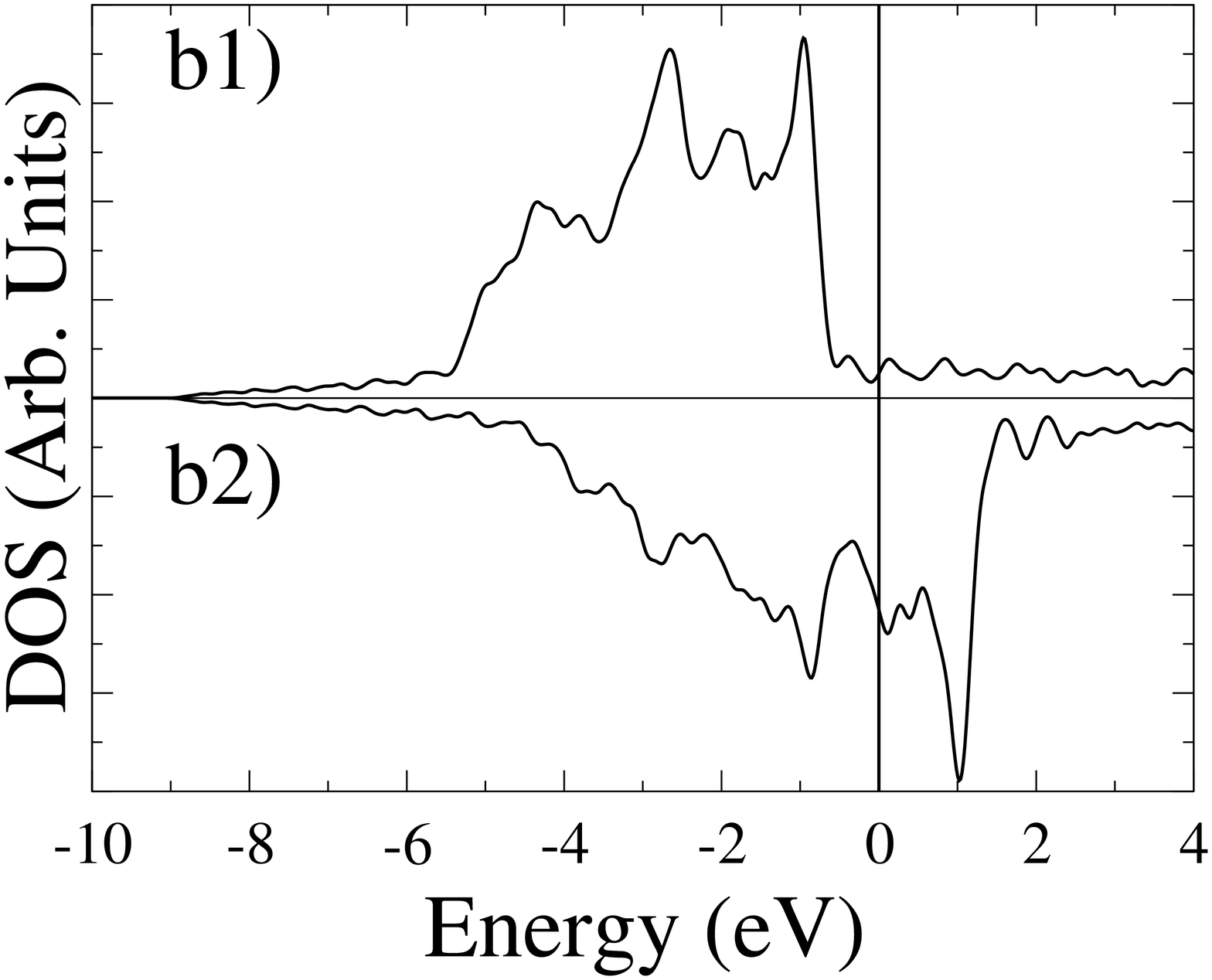}
\vspace{0.5cm}
\includegraphics[width=5.5cm,clip=true]{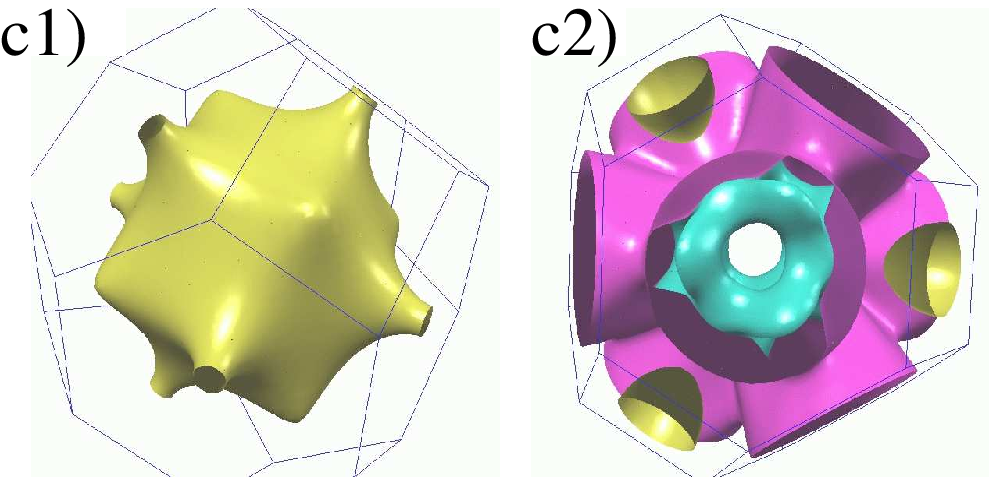}
\end{center}
\caption{a) Band structure, b) density of states, and c) Fermi surface for fcc Co. The figures a1), b1) and c1) refers to the majority spin electrons, while a2), b2) and c2) to the minority. The pictures a) and b) have been 
obtained with density functional theory using the code SIESTA \cite{siesta}, and c) from an
$spd$ tight-binding Hamiltonian \cite{Fermi_per}.}
\label{fig1}
\end{figure}

In addition to the formation of a net magnetic moment a consequence
of the bandstructure spin-splitting is that the Fermi surface for the two spin directions
is rather different. This difference is more pronounced in the case of strong ferromagnet, 
where only one of the two spin-split $d$ manifolds is fully occupied (majority band) while 
the other has some fractional occupation (minority band). An example of this situation is fcc Co 
(the high temperature phase), whose electronic structure is presented in figure \ref{fig1}. 

The main feature of the bandstructure of Co (and indeed of the other 3$d$ magnetic transition
metals) is the presence of a broad (delocalized) and only weakly spin-split $s$ band, and of
a narrow (localized) and largely spin-split $d$ band. The former has almost free electron-like character
for energies both below (note the parabolic behavior of the bands around $\Gamma$ for $E\sim$~-9~eV) 
and above $E_\mathrm{F}$. In contrast
the latter is only about 5~eV wide and cuts close to the Fermi level. Because of the spin-splitting
of the $d$ manifold, $\Delta$, the majority spin band has an almost spherical Fermi surface 
(see figure \ref{fig1}c1), while the minority one has a rather complicated structure, mostly 
arising from the $d$ manifold (see figure \ref{fig1}c2). The different structure of the Fermi 
surface for majority and minority spins and the fact that these differences arise from a different
orbital character are the main ingredients for understanding the transport properties of magnetic 
transition metal heterostructures.

Importantly the effects of alloying can be understood in the context of a rigid-band model, 
by shifting the Fermi level according to the valence of the dopant \cite{Slater,Pauling}.
Finally it is worth mentioning that there are materials that at the Fermi level present a 
finite DOS for one spin specie and a gap for the other. These are known as half-metals 
\cite{ss_2} and are probably among the best candidates as materials for future magneto-electronics 
devices.

\subsection{Spin valves}

The prototype of all spin devices is the spin-valve. This is  formed by two magnetic
layers (normally transition metals) separated by a non-magnetic spacer (either
metal or insulator). One of the two magnetic layers is free to rotate in tiny magnetic fields,
while the other usually is pinned by exchange coupling with a antiferromagnet or by strong
magnetic anisotropy. The current passing through a spin-valve depends over the mutual
orientation of the two magnetic layers and it is typically higher for a parallel alignment (PA) than
for an antiparallel (AA). Thus a spin-valve behaves essentially as a spin polarizer/analyzer 
device. The quantity that defines the effectiveness of the spin-filtering effect is the GMR ratio
$R_\mathrm{GMR}$ defined as (``optimistic definition'')
\begin{equation}
R_\mathrm{GMR}=\frac{I_\mathrm{PA}-I_\mathrm{AA}}{I_\mathrm{AA}}\:.
\end{equation}
An alternative definition (``pessimistic definition'') using $I_\mathrm{PA}+I_\mathrm{AA}$ as 
normalization is sometime used.

\begin{figure}[htb]
\begin{center}
\includegraphics[width=11.0cm,clip=true,angle=-90]{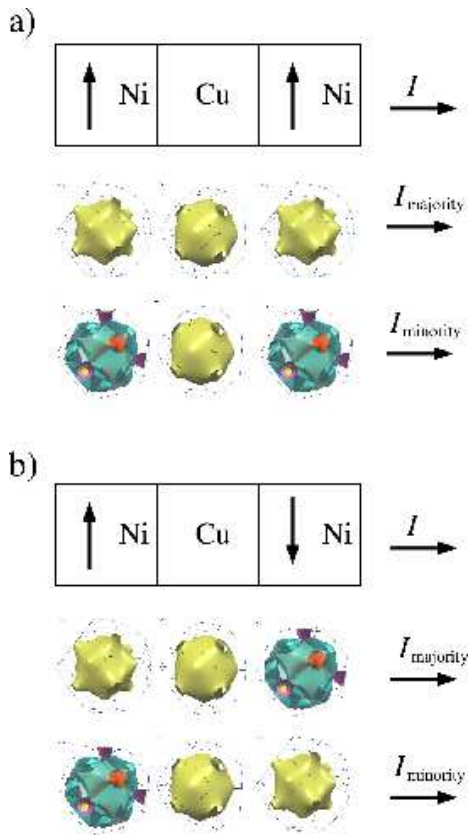}
\end{center}
\caption{Magnetoresistance mechanism in a Ni/Cu/Ni spin valve in the two spin fluid approximation.
In the parallel case a) the resistance of the majority spin channel is low since there is good match of
the majority Fermi surfaces across the entire device. In contrast 
in the antiparallel state b) the alignment of the Fermi surface is such to have one high resistance interface
for both the spin channels. This is given by the interface between the minority spin band of Ni and that of Cu. 
The formation of high resistance channels for both spins in the AA is responsible for the GMR effect.}
\label{fig2}
\end{figure}

An intuitive understanding of the spin-filtering produced by a spin-valve can be obtained by 
looking at the Fermi level lineup of the materials forming the device. Let us consider for
example a Ni/Cu/Ni spin-valve (see figure \ref{fig2}), and assume that the two spin-bands
do not mix. This is the two spin fluid approximation, which is valid in the case of weak spin-orbit
scattering and collinear magnetism \cite{2current_mott}.

In the AA a spin electron propagates in the majority spin band in one layer and in the minority band
in the other. Consequently electrons always travel across the Fermi surfaces of Cu and of both the 
spin-bands of Ni. In contrast in the PA the two 
spin currents are rather different. The majority current is made from electrons that have traveled within 
the Fermi surfaces of Cu and that of the majority spin of Ni, while the down spin current 
from electrons that have traveled within the Fermi surfaces of Cu and that of the minority spin of Ni. 
This leads to two different current paths for the AA and the PA. Since the two spin currents add to form the total current
and since generally the resistances of majority and minority electrons are different in a magnetic transition metal, 
the current passing through the PA and AA configurations are different. Importantly the larger is the mismatch 
between the two spin currents, the larger in the GMR ratio.

\subsection{Spin polarization of a device}

An important question is how to quantify the relative difference between the two spin currents in a magnetic device
and how to relate this properties to the elementary electronic structure of the materials forming the
device. We then define the spin-polarization $P$ of a material/device as
\begin{equation}
P=\frac{I_\uparrow-I_\downarrow}{I_\uparrow+I_\downarrow}\:,
\label{Pdef}
\end{equation}
where $I_\sigma$ is the spin-$\sigma$ contribution to the current. $I_\sigma$ and $P$ are 
not directly observable and must be calculated or inferred from indirect measurements. 
Unfortunately the way to relate the spin-current $I_\sigma$ to the electronic structure of a material 
is not uniquely defined and depends on the particular experiment carried out. 

As brilliantly pointed out by Mazin \cite{Mazin_PRL}, the relation between the 
spin-polarization of a magnetic material and its electronic structure depends critically 
on the transport regime that one is considering (ballistic, diffusive, tunneling ..). 
As a first approximation the current $I$ is simply proportional to $N_\mathrm{F} v_\mathrm{F}^n$,
where $N_\mathrm{F}$  and $v_\mathrm{F}$ are the DOS at the Fermi level and the Fermi 
velocity respectively. Different transport regimes weight the contribution of the Fermi velocity differently,
and one has $n=2$ for diffusive transport, $n=1$ for ballistic transport and $n=0$ for tunneling.

Therefore the spin-polarization $P$ becomes
\begin{equation}
P_{n}=\frac{N_\mathrm{F}^\uparrow (v_\mathrm{F}^{\uparrow})^n-
N_\mathrm{F}^\downarrow (v_\mathrm{F}^{\downarrow})^n}
{N_\mathrm{F}^\uparrow (v_\mathrm{F}^{\uparrow})^n+
N_\mathrm{F}^\downarrow (v_\mathrm{F}^{\downarrow})^n}\:.
\label{Pn}
\end{equation}
Typical values of $P_{n}$ for several magnetic metals are reported in table \ref{Table1}.
\begin{table}[ht]
\begin{center}
\begin{tabular}{lcccc}
\hline
\hline \\[-0.2cm]
  & $P_{n}$   (\%) & $n$=2 & $n$=1 & $n$=0  \\[0.1cm]
\hline \\[-0.2cm]
Fe      		&	& 20 & 30 & 60 \\
Ni      		&	& 0 & -49 & -82 \\
CrO$_2$ 		&	& 100 & 100 & 100 \\
La$_{0.67}$Ca$_{0.33}$MnO$_3$ &	& 92 & 76 & 36 \\
Tl$_2$Mn$_2$O$_7$	&	& -71 & -5 & 66 \\[0.2cm]
\hline
\hline
\end{tabular}
\end{center}
\caption{Spin-polarization of typical magnetic metals according to the various definitions given
in the text. The data are taken from literature as follows: Ni and Fe \cite{Mazin_PRL}, 
CrO$_2$ \cite{CrO}, La$_{0.67}$Ca$_{0.33}$MnO$_3$ \cite{LCMO} and Tl$_2$Mn$_2$O$_7$
\cite{TMO}.}
\label{Table1}
\end{table}

Importantly the spin-polarization of a device can be different from that of the materials forming
it. This is connected to the fact that the bonding at the interface between two different materials
can be strongly spin-selective. For instance if the bonding between two materials has mainly
$s$-character, then one expects strong scattering for $d$-like electrons. As a consequence 
the spin-polarization of the current will be determined by the spin-polarization
of the almost free $s$-electrons. This is usually much smaller than that of the $d$-electrons 
and it may even have the opposite sign. For instance it is demonstrated that the
sign of the magnetoresistance of a magnetic tunneling junction can be altered by simply replacing 
the insulator forming the barrier \cite{deteresa}.

\section{Molecular-Electronics}
\subsection{Electrons transport through molecules}
\begin{figure}[htb]
\begin{center}
\includegraphics[width=3.0cm,clip=true,angle=-90]{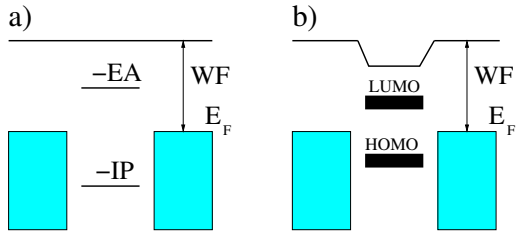}
\end{center}
\caption{Energy level lineup between a molecule and two current voltage probes. In the weak coupling 
limit a) the molecule is characterized by the ionization potential (IP) and the electron affinity (EA), which line
up with the metal Fermi energy (WF is the work function of the contact). In the case of strong coupling b) 
between the molecule and the leads the molecular levels shift and broaden. 
It is then more appropriate to discuss transport in terms of HOMO and LUMO states.}
\label{fig3}
\end{figure}

Transport through molecules and in general through low dimensional objects is somehow different 
than that in standard metals or semiconductors. This is substantially due to the collapse of the
Fermi surface into a single energy level (the highest occupied molecular orbital - HOMO). The nature
and lineup of the HOMO with the Fermi energy of the current/voltage probes determine most of the transport properties. Let us consider the simple case of a two probe device. Following a simple model proposed by 
Datta \cite{datta} the typical energy level lineup is schematically presented in figure \ref{fig3}.

In absence of any coupling (figure \ref{fig3}a) both the energy levels of the molecule and the Fermi level
of the electrodes will align with a common vacuum level. In this case the system is characterized
by the work function of the electrodes and both the ionization potential (IP) and the electron affinity (EA)
of the molecule. In this setup the molecule can exchange electrons with the electrodes only if the
typical temperature is comparable to either  IP-WF or WF-EA, a condition which is normally not
satisfied. This guarantees local charge neutrality of the whole system and integer occupation of the molecule.

In contrast, the interaction between the molecular levels and the extended wave-functions of the metallic contacts
has the effect of broadening and shifting the molecular levels. In the extreme limit of large coupling 
extended states spanning through the entire system (electrode plus molecule) can develop and 
the molecular device will behave as a good metal. In this limit the molecular levels cannot
be associated any longer to the elementary removal energies of the isolated molecule and a description
in terms of fractionally occupied HOMO and LUMO (lowest unoccupied molecular orbital) is more
appropriate. 

The transition from integer to fractional occupation of the molecule somehow depends on the
typical molecular charging energy (say the EA) compared to the hopping integral $\Gamma$ between
the molecule and the contacts ($\Gamma/\hbar$ is the escape rate from the molecule to the contacts).
One has integer occupation if EA~$\gg\Gamma$ and metallic-like behaviour when $\Gamma\gg$~EA.
This is essentially the same physics leading to Mott metal-insulator transition in solid state.
An electronic structure theory capable of exploring on the same footing all the intermediate 
situations between the strong and weak coupling limit is still not available unless at prohibitive computational
costs \cite{SmeagolSIC}.

The effect of an applied bias $V$ is that of shifting the chemical potentials of the two current/voltage
probes relative to each other by $eV$, with $e$ the electronic charge. As a rule of thumbs current
will flow whenever a molecular level (either the HOMO or the LUMO) is positioned within such 
a bias window. The appearing of molecular levels in the bias window when the potential is increased 
typically leads to changes in the slope of the $I$-$V$ characteristics, in steps in the differential conductance 
$\mathrm{d}I/\mathrm{d}V(V)$ and in peaks in its derivative $\mathrm{d}^2I/\mathrm{d}V^2(V)$. 
This means that fingerprints of the molecular spectrum can be found in the measurement of its electrical 
properties. The same is true for the molecular elementary excitations, and peaks in $\mathrm{d}^2I/\mathrm{d}V^2(V)$
can be found in correspondence of the energy of relevant phonon modes \cite{IETS}.

\subsection{The bonding with the contacts}
\begin{figure}[htb]
\begin{center}
\includegraphics[width=5.0cm,clip=true,angle=-90]{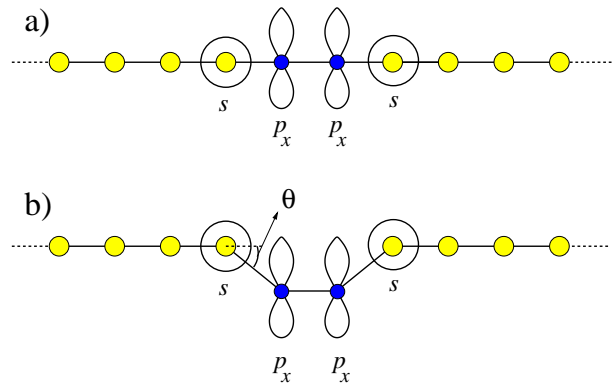}
\end{center}
\caption{Au atomic chain sandwiching a $\pi$-bonded molecule (say an S$_2$ molecule). 
a) the molecule is aligned with the Au chain and the transmission is suppressed because the 
matrix element $\langle s|H|p_x\rangle$ vanishes. In contrast when the molecule forms some 
angle $\theta$ with the Au chain b) then a component of the hopping integral along the bond 
develops and current can flow.}
\label{fig4}
\end{figure}

One of the fundamental aspects of molecular electronics is that the bonding between a molecule
and the current/voltage probes can be engineered to a degree usually superior to that 
achievable in conventional inorganic heterostructures. This can dramatically change the 
current flowing through a device. Consider for instance the simple case of an atomic gold chain,
described by $s$ orbitals only, sandwiching a $\pi$-bonded molecule (see figure \ref{fig4}).

Let us assume that the relevant molecular state (the one close to the Fermi level of the gold chain) is 
formed by $p_x$ orbitals, i.e. those perpendicular to the chain axis. When the molecule is positioned 
along the axis of the chain the hopping integral between the molecule and the chain $\langle s|H|p_x\rangle$ 
vanishes regardless of the separation between the two. This is simply the result of 
the particular symmetry of the problem since $s$ and $p_x$ orbitals do not share the same
angular momentum about the bond axis \cite{sutton}. As a consequence the current is identically zero.

\begin{figure}[htb]
\begin{center}
\includegraphics[width=6.5cm,clip=true]{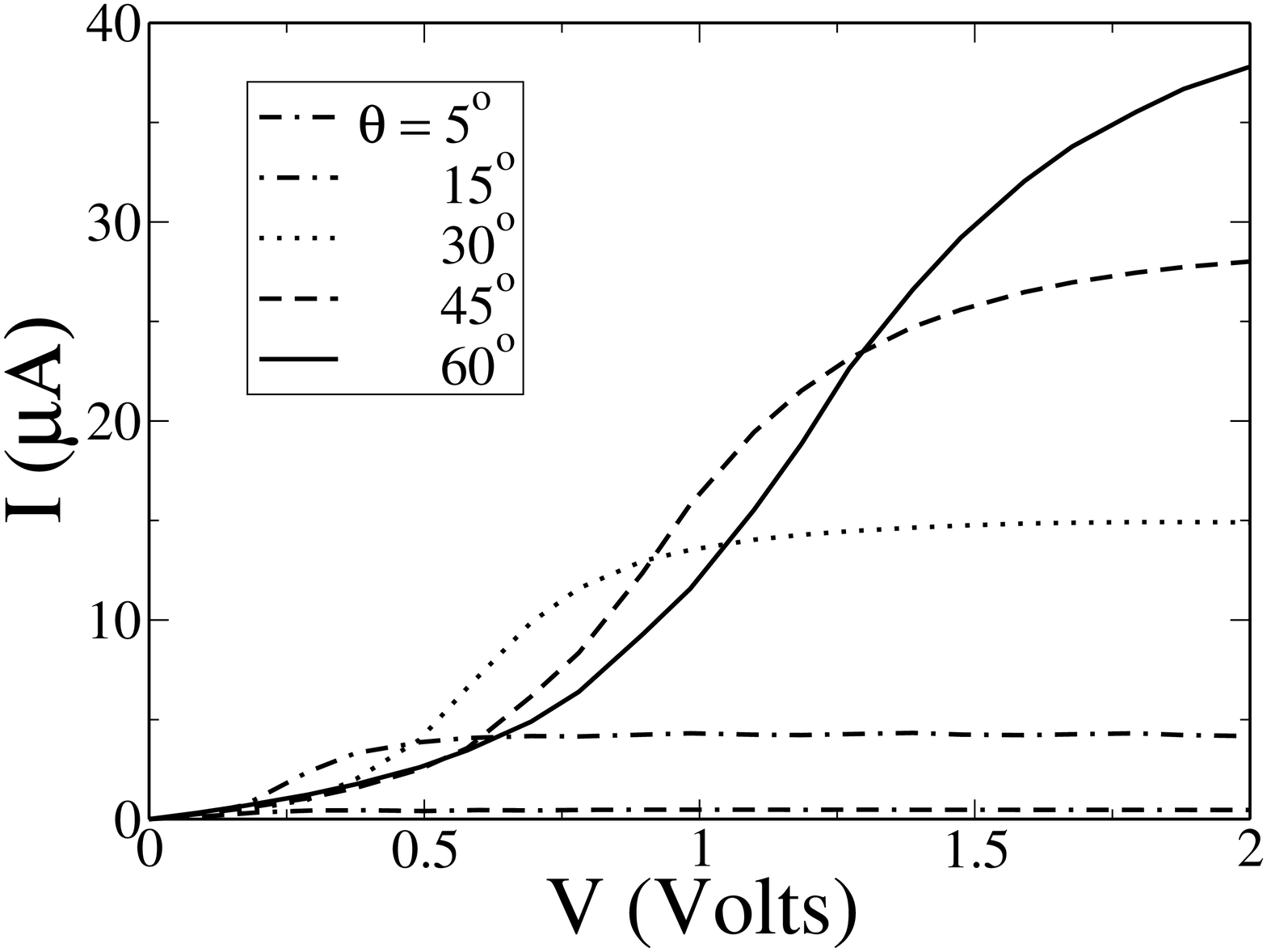}
\end{center}
\caption{$I$-$V$ characteristics for a S$_2$ molecule sandwiched between two semi-infinite
gold chains (see figure \ref{fig4}). The curves are calculated using the tight-binding based 
non-equilibrium Green function method of reference \cite{SmeagolSIC} with the following parameters:
$\epsilon_\mathrm{Au}$=-5.9~eV, $\epsilon_\mathrm{S}$=-6.15eV~eV, $\gamma_{ss\sigma}$=-3.0~eV,
$\gamma_{sp\sigma}$=1.52~eV, $\gamma_{pp\pi}$=-0.98~eV, $U_\mathrm{Au}$=-6.7~eV and 
$U_\mathrm{S}$=-6.15~eV. $\epsilon$ is the on-site energy, $\gamma$ hopping
integral and $U$ the charging energy.}
\label{fig5}
\end{figure}

In contrast if the molecule is not coaxial to the chain (figure \ref{fig4}b), then there is a component 
of the $p_x$ orbital along the bond axis and the hopping integral becomes $\gamma_{sp\sigma}\:\sin\theta$,
where $\gamma_{sp\sigma}$ is the $sp\sigma$ hopping integral and $\theta$ the bond angle.
This dependence of the bonding on the bond orientation may have dramatic consequences on the 
$I$-$V$ characteristics. In figure \ref{fig5} we present the $I$-$V$ curve for the system of
figure \ref{fig4} calculated with a self-consistent tight-binding model where it is assumed a
linear dependence of the on-site energies over the orbital occupation (see reference [\onlinecite{SmeagolSIC}]
for details).

From figure \ref{fig5} it is clear that the $I$-$V$ characteristics of a molecule can be largely 
engineered by simply changing the details of the bonding with the electrodes. This is considerably 
more complicated in extended interfaces (for instance between two metals), since disorder,
interdiffusion and roughness have the effect of averaging out the atomistic details of the bonding.
It is important to remark that even when molecular layers are grown a good level of tuning of the
bonding properties still exists. For instance the bonding site and the bond angle usually depend 
on the layer density (coverage) \cite{coverage}, and these can be further tuned by changing 
the end groups. 

\subsection{Why spins and molecules?}
What are the advantages of using molecules instead of inorganic materials for performing 
spin-physics? These are essentially two. On the one hand there are intrinsic molecular
properties and in particular the weak spin-orbit and hyperfine interactions.
On the other hand there are the properties connected to the formation of interfaces between 
magnetic metals and molecules. We will return to the interfacial properties in the next sections,
here we focus our attention only on the intrinsic aspects.

Spin-orbit interaction is a relativistic effect 
which couples the electron spin $\vec{S}$ with its angular momentum $\vec{L}$. 
The spin-orbit Hamiltonian in general can
be written as $H_\mathrm{SO}=V_\mathrm{SO}\vec{S}\cdot\vec{L}$, where $V_\mathrm{SO}(\vec{r})$
is a term which contains the gradient of the electrostatic potential. Although it is rather intuitive to
realize that the strength of this interaction grows with the atomic number $Z$ (it is proportional to $Z^4$), its actual value in the solid
state depends on various factors such as the crystal symmetry and the material composition. Importantly
the spin-orbit effect is responsible for spin-precession and the loss of spin-coherence. In organic materials
usually the spin-orbit interaction is rather small. This is mostly due to the small atomic number of carbon.
In table \ref{tab2} we compare the spin-orbit splitting $\Delta_\mathrm{SO}$ of the valence band of several 
semiconductors \cite{SOmiguel} with that of carbon diamond \cite{SOCarbon}.
\begin{table}[ht]
\begin{center}
\begin{tabular}{lc}
\hline
\hline \\[-0.2cm]
  & $\Delta_\mathrm{SO}$   (meV)   \\[0.1cm]
\hline \\[-0.2cm]
Si 		& 44 \\
Ge		& 290 \\
GaAs		& 340 \\
AlAs		& 280 \\
InAs		& 380 \\
GaP		& 80 \\
InP		& 111 \\
GaSb		& 750 \\
AlSb		& 670 \\
InSb		& 980 \\
C		& 13  \\[0.2cm]
\hline
\hline
\end{tabular}
\end{center}
\caption{Valence band spin-orbit splitting for various semiconductors.}
\label{tab2}
\end{table}
The table shows that in carbon the spin-split of the valence band is approximately one
order of magnitude smaller than in ordinary III-V or group IV semiconductors and one should expect
a considerably longer spin-lifetime \cite{Aws1,Aws2}. 

Another important interaction, which generally leads to spin-decoherence, is the hyperfine interaction
between electron and nuclear spins. This has the form
\begin{equation}
H_\mathrm{hyp}=A_\mathrm{hyp}\:\vec{s}\cdot\vec{S}_\mathrm{N}\:,
\end{equation}
where $\vec{s}$ and $\vec{S}_\mathrm{N}$ are respectively the electronic and nuclear spin,
and $A_\mathrm{hyp}$ is the hyperfine coupling strength.
Similarly to the case of spin-orbit, also hyperfine 
interaction is a source of spin de-coherence \cite{Loss}, since the random flipping 
of a nuclear spin can cause
that of an electron spin. However, in III-V semiconductors hyperfine 
interaction was also proved to be a tool for controlling nuclear spins via optically polarized 
electron spins \cite{AwsNMR}. In organic materials usually the hyperfine interaction is weak.
The main reason for this is that most of the molecules used for spin-transport are 
$\pi$-conjugate molecules where the transport is mostly through molecular states localized 
over the carbon atoms. Carbon, in its most abundant isotopic form, $^{12}$C,
has nuclear spin $S_\mathrm{N}$=0, and therefore is not hyperfine active. 
Moreover the $\pi$-states are usually delocalized and $H_\mathrm{hyp}$
can be anyway rather small. In table
\ref{tab3} we report the value of the nuclear spin for various atomic species with their
relative isotopic abundance.
\begin{table}[ht]
\begin{center}
\begin{tabular}{lccc}
\hline
\hline \\[-0.2cm]
Isotope & IA (\%) & $S_\mathrm{N}$ &     \\[0.1cm]
\hline \\[-0.2cm]
$^{1}$H 		& 99.98 & 1/2   \\
$^{2}$H 		& 0.02 & 1   \\
$^{12}$C 		& 98.93 & 0   \\
$^{13}$C 		& 1.1 & 1/2   \\
$^{14}$N 		& 99.632 & 2   \\
$^{15}$N 		& 0.368 & 1/2   \\
$^{16}$O 		& 99.757 & 0   \\
$^{18}$O 		& 0.205 & 0   \\
$^{19}$F 		& 100 & 1/2   \\
$^{69}$Ga 		& 60.108 & 3/2    \\
$^{71}$Ga 		& 39.892 & 3/2    \\
$^{75}$As 		& 100 & 3/2    \\
$^{28}$Si 		& 92.2297 & 0    \\
$^{29}$Si 		& 4.6832 & 1/2    \\
$^{30}$Si 		& 3.0872 & 0    \\ \\[0.1cm]
\hline
\hline
\end{tabular}
\end{center}
\caption{Nuclear spin for elements present in typical organic molecules and in both Si and GaAs. 
Here we report the nuclear spin for the most abundant isotopes, together with their relative 
isotopic abundance (IA).}
\label{tab3}
\end{table}

Estimates of the spin-lifetime of organic materials from transport experiments are at the 
moment only a few. Moreover in most cases these are extracted from spin-valves measurements
by fitting to the Jullier's formula \cite{Jullier}. This procedures does not distinguish the source
of spin-flip, which may not be located inside the molecule, but at the interface with the magnetic
electrodes. Therefore these measurements are likely to offer a conservative estimate of the
spin-lifetime. Nevertheless the values of the spin diffusion length reported in the literature are 
rather encouraging for carbon nanotubes (130nm) \cite{bruce}, polymers 
(200nm) \cite{Dediu}, or Alq$_3$ molecules (5nm) \cite{Alq3}.

Finally a conclusive note needs to be made on conducting polymers. These are extremely attractive
materials since their electrical conductivity can be changed by over twelve orders of magnitude 
with the chemical or electrochemical introduction of various counterions \cite{SSH}. Spin-dynamics
in polymers has been extensively studies with EPS spectroscopy \cite{EPR} and it 
is largely dominated by the presence of paramagnetic centers in the form of free radicals, 
ion-radicals, molecules in triplet states and transition metal complexes. 
Even more interesting is the fact that the elementary 
excitations leading to electron transport are not band-like but usually involve lattice vibrations, and
most importantly some of them are spin-polarized. 

For instance in the {\it trans}-isomer of 
polycetylene the Su-Schrieffer-Heeger theory \cite{SSH} predicts a soliton-like transport
mechanism. This has a peculiar spin-charge relationship, since a neutral soliton corresponds
to a radical with spin 1/2, while both negatively and positively charge solitons are spinless and
diamagnetic. The study of spin-transport in devices made by transition metals in contact with such
polymers is potentially extremely interesting. This is because of the proximity of two fundamentally 
different ground states, one electronically correlated (the magnetic material), and one
with strong correlation between electronic and vibrational degrees of freedom (the polymer). 
The investigation, both experimentally and theoretically, of these combined systems is in its
infancy \cite{Bishop}.

\section{Quantitative Transport Theory}
\subsection{Simple Model}

Modern theory of quantum transport is based on scattering theory
in conjunction with accurate electronic structure methods. Although this approach has
recently come to question, in particular in the case of electrons interacting beyond the 
mean-field level \cite{kieron}, it still remains the most versatile and scalable available.
In addition its foundations are extremely intuitive and simple. Let us start our
discussion by presenting a simple model, first introduced by Datta \cite{datta}, which already contains all
the elements of a more formal and accurate theory.

Consider a given molecule attach to two current/voltage probes kept at two different
chemical potentials $\mu_\alpha$, with $\alpha$=L (left), R (right). The leads are assumed 
featureless, which is with a constant DOS. The molecule is described
by an energy level $\epsilon$ (say the HOMO), coupled to the current/voltage probes
by the hopping integrals $t_\alpha$ ($\alpha$=L, R). The density of states associated to such a state is
\begin{equation}
D_\epsilon(E)=\frac{t/2\pi}{(E-\epsilon)^2+(t/2)^2}\;,
\label{Lor}
\end{equation}
where the level lifetime $\hbar/t$ is determined by the coupling with the leads only $t=t_\mathrm{L}+t_\mathrm{R}$. Both the current $I$ flowing through the molecular state and
state occupation $N$ can be determined by balancing the in-going and out-going fluxes
and read
\begin{equation}
I=\frac{e}{\hbar}\int_{-\infty}^{+\infty}\mathrm{d}E\:D_\epsilon(E)
\frac{t_\mathrm{L}t_\mathrm{R}}{t}[f_\mathrm{L}(E)-f_\mathrm{R}(E)]\;,
\label{Nscb}
\end{equation}
and
\begin{equation}
N=\int_{-\infty}^{+\infty}\mathrm{d}E\:D_\epsilon(E)
\frac{t_\mathrm{L}f_\mathrm{L}(E)+t_\mathrm{R}f_\mathrm{R}(E)}{t}\;,
\label{Nsmb}
\end{equation}
where $f_\alpha(E)=\frac{1}{1+\mathrm{e}^{(\epsilon-\mu_\alpha)/k_\mathrm{B}T}}$ is the
Fermi distribution of the contact $\alpha$. From the equations above it is clear that current will
flow only (at least for small broadening and low temperature) if the energy level is
in between the chemical potentials of the two leads. With no external bias these are
identical, but when a potential $V$ is applied then $\mu_\mathrm{L}-\mu_\mathrm{R}=eV$
and current will flow. 

In general $\epsilon$ depends on the details of the electronic structure of the system. As a simple
approximation we may assume it is a function of the level occupation
only $\epsilon=\epsilon(N)$. This suggests a simple self-consistent procedure
where $\epsilon(N)$ and the equation (\ref{Nsmb}) are solved iteratively before the current
is evaluated with the (\ref{Nscb}).

Spin can be easily introduced in this simple model in the two spin fluid approximation. 
Since the two spin channels do not mix the equations (\ref{Nscb}) and (\ref{Nsmb}) can be
replaced by two pairs of equations for the spin-resolved molecular level occupation $N^\sigma$
and the spin-current $I_\sigma$. In a similar way the molecular level can also be spin-polarized
$\epsilon\rightarrow\epsilon^\sigma$ and it may depend of the spin density instead of the
density only $\epsilon^\sigma(N^\uparrow,N^\downarrow)$. Importantly in the case where the current/voltage
probes are ferromagnetic spin-degeneracy is lifted by introducing spin-dependent hopping
integrals $t_\alpha^\sigma$ between the leads and the molecule. These however capture
the fact that majority and minority electrons couple to the molecule in a different way, but not
that the DOS for the different spin directions in a ferromagnet is different.
For this last feature a more detailed description of the electronic structure of the leads is needed.

\subsection{{\it Ab initio} methods}

The non-equilibrium Green's function (NEGF) method is by far the most used among all the 
quantum transport schemes. Although it is based on very rigorous ground \cite{haug} it can
be understood as the natural extension of the toy-model discussed in the previous section. 
The general idea is to divide a two probe device into three regions: two current/voltage
probes and a scattering region. The criterion for this fragmentation is that the scattering region
is the portion of the device where the potential drops. Its boundaries are defined by the condition 
that the charge density matches exactly that of the bulk material forming the leads (see reference
\cite{Smeagol1} for a more detailed description). 

Let us assume that the problem can be formulated over some sort of localized basis
set, and therefore the Hamiltonian for the whole system (scattering region plus leads) is simply 
an infinite hermitian matrix. The central quantity, which replaces the simple density of states
$D_\epsilon(E)$, is the non-equilibrium Green's function for the scattering region $G(E)$
\begin{equation}
G(E)=\lim_{\eta\rightarrow 0}[(E+i\eta)-H_\mathrm{S}-\Sigma_\mathrm{L}-\Sigma_\mathrm{R}]^{-1}\:.
\label{negfmx}
\end{equation}
This is the Green's function associated to the Hamiltonian $H_\mathrm{S}+\Sigma_\mathrm{L}+\Sigma_\mathrm{R}$, which is composed
by the Hamiltonian of the scattering region $H_\mathrm{S}$ and the self-energies
$\Sigma_\mathrm{L}$ and $\Sigma_\mathrm{R}$. 

Note that if one assumes that $H_\mathrm{S}$, $\Sigma_\mathrm{L}$ and 
$\Sigma_\mathrm{R}$ are just C-numbers, then $i[G(E)-G^*(E)]$ is
$D_\epsilon(E)$ for $\epsilon\rightarrow H_\mathrm{S}$ (real) and $t_\alpha/2\rightarrow\Sigma_\alpha$
(purely imaginary). Thus the self-energies can be associated with
the interaction between the scattering region and the current/voltage probes. In practice they
can be written as $\Sigma_\mathrm{L}=H^\dagger_{\mathrm{LS}}g_\mathrm{L}H_\mathrm{LS}$
and $\Sigma_\mathrm{R}=H_{\mathrm{RS}}g_\mathrm{R}H^\dagger_\mathrm{RS}$,
with $H_{\alpha\mathrm{S}}$ the coupling matrix between the leads $\alpha$ and the
scattering region. $g_\alpha$ are the surface Green's functions for the leads, i.e. the Green's
function for a semi-infinite lead evaluated at the termination plane \cite{rgf}. Thus the 
self-energies contain information on both the coupling between the scattering region and 
the electrodes and on the electronic structure of the leads themselves.

The crucial point is that both the two-probe current $I$ and the density matrix associated
to the scattering region $\rho$ can be obtained from the Green's function $G(E)$
\begin{equation}
I=\frac{e}{h}\int_{-\infty}^{+\infty}\mathrm{d}E\:\mathrm{Tr}[G\Gamma_\mathrm{L}
G^\dagger\Gamma_\mathrm{R}][f_\mathrm{L}(E)-f_\mathrm{R}(E)]\;,
\label{CurrNEGF}
\end{equation}
and 
\begin{equation}
\rho=\frac{1}{2\pi}\int dE\:
G[\Gamma_\mathrm{L}f_\mathrm{L}+\Gamma_\mathrm{R}f_\mathrm{R}(E)]
G^\dagger\:,
\label{denmatrix}
\end{equation}
where $G=G(E)$, $\Gamma=\Gamma(E)$ we have now introduced the broadening matrices
\begin{equation}
\Gamma_\alpha=i[\Sigma_\alpha-\Sigma_\alpha^\dagger]\:.
\label{broadmat}
\end{equation}

Equations (\ref{negfmx}) and (\ref{CurrNEGF}) allow us to calculate the current once the
Hamiltonian for the scattering region and the self-energies are given. Unfortunately these are
not known {\it a priori}, since they require the evaluation of the electronic structure of an infinite
non-periodic system. However their calculation do not require the actual solution of this open
system and the notion of locality (the same that allows us to separate the leads from the scattering
region) can be efficiently used. The self-energies in fact involve only bulk quantities, i.e. quantities
which can be evaluated from a bulk calculation for a periodic system. In addition a self-consistent
scheme can be designed for evaluating $H_\mathrm{S}$. The crucial point here is to assume
that the whole electronic structure can be described by a single-particle theory, in such a way that
$H_\mathrm{S}$ depends only on the density matrix $H_\mathrm{S}=H_\mathrm{S}[\rho]$. 
This is for instance the case of DFT or Hartree-Foch methods. Hence
$H_\mathrm{S}[\rho]$ and the equations (\ref{negfmx}) and (\ref{denmatrix}) can be iterated self-consistently
until convergence, and then the current can be evaluated with equation (\ref{CurrNEGF}). 
This scheme, with differences concerning the specific numerical implementation, is used by most of the 
DFT-based NEGF packages available at 
present \cite{Smeagol1,Smeagol2,NEGF1,NEGF2,NEGF3,NEGF4,NEGF5}.

Also in this case the addition of spin polarization does not bring any fundamental changes in the
formalism. In the simple case of collinear-spins, then all the quantities introduced (Green's 
functions, charge density, Hamiltonian, self-energies ...) becomes block diagonal matrices 
in spin space. The Hamiltonian
now depends on both the charge density and the magnetization as in standard spin-polarized
mean field electronic structure methods. Importantly the current is still carried in parallel by the two spin 
species. In the more complicated case of non-collinear spin (or if spin-orbit is present) then the off-diagonal blocks to not vanish and the current cannot be broken down into the majority and minority
contributions.

Note that computationally the introduction of magnetism usually makes the calculation more
complicated. First one needs to consider matrices larger than the unpolarized case. Secondly and
most importantly the Hamiltonian matrix becomes considerably more sparse. In a ferromagnet in fact,
for instance in a transition metal, the delocalized $s$-electrons responsible for most of the electron transport coexist with the tightly bound $d$-electrons, which provide the local magnetic moment.
This means that most of the matrix elements between $d$ orbitals located on atoms far from each other
vanish, making the Hamiltonian more sparse than that of standard free-electron like metals 
(for instance Au). In the case of sparse matrices standard recursive methods for evaluating the 
self-energies become numerically unstable if not prohibitive, and more sophisticated schemes are
needed \cite{Smeagol1,Smeagol2}.

\section{Molecular Spin-valves}

As pointed out in the introduction spin-valves are the prototypical spin-devices and therefore
are the most studied architectures for spin-transport through molecules. The first prediction of
a GMR-like effect in molecules is from Emberly and Kirczenow, who investigated the transport
through a 1,4-benzeneÐdithiolate molecule attached to Ni electrodes. Because of the problems
of dealing with magnetic current/voltage probes mentioned in the previous section, this work
and most of the early calculations \cite{emberley} were limited to empirical models for the electronic
structure.

\begin{figure}[htb]
\begin{center}
\includegraphics[width=8.5cm,clip=true]{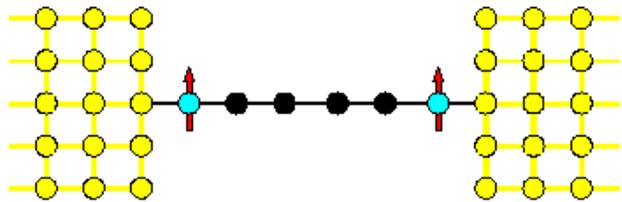}
\end{center}
\caption{Setup for early molecular spin-valve calculations. A molecule (black circles)
is sandwiched between two magnetic ions (blue circles), which are then contacted by two 
non-magnetic current/voltage electrodes. The red arrows indicate the local magnetic moments.}
\label{fig6}
\end{figure}

A first step in the direction of using {\it ab initio} electronic structure methods was suggested by
Pati and co-workers \cite{Pati1,Pati2,Pati3}, who considered a setup in which a given molecule is sandwiched 
between two magnetic ions, or clusters, which then are 
attached to non-magnetic electrodes (see figure \ref{fig6}). 

This setup removes the problems connected with constructing the self-energies for magnetic 
materials and a GMR-like effect is produced by the magnetic ions, which in turn are exchanged
coupled through the molecule. Clearly the scheme is highly idealized. It essentially describes
the transport through a magnetic molecule (molecule plus magnetic ions) from non-magnetic leads,
and not that of a magnetic spin-valve. Importantly it does not account for both the spin-polarized 
DOS of the contacts and the accurate orbital character of the bonding between the molecule and the 
magnetic surfaces. 

An interesting alternative, that still avoids the construction of magnetic leads,
was proposed by Wei and co-workers \cite{GuoAl}, who used Al leads locally immersed in a
magnetic field. In this case the spin-polarization of the leads is achieved through simple Zeeman 
splitting and the direction of the local magnetic fields at the two contacts replaces that of the magnetic 
moments of a ferromagnet. In this case the DOS of the leads is spin-polarized, however the orbital
nature of the bonding is identical for both majority and minority spins. 
Importantly most of these early calculations were ahead of experiments and largely inspired them.

\subsection{Metallic and tunneling junctions}

The study of molecular transport with ferromagnetic contacts from first principles has a young
history since only recently algorithms stable enough to deal with extremely sparse matrices
and thus with magnetic leads were made available \cite{Smeagol1,Smeagol2}. 
Here we describe the results obtained with 
the code {\it Smeagol}~\cite{Smeagol1,Smeagol2} for both insulating and tunneling molecules.

\begin{figure}[htb]
\begin{center}
\includegraphics[width=5.5cm,clip=true,angle=-90]{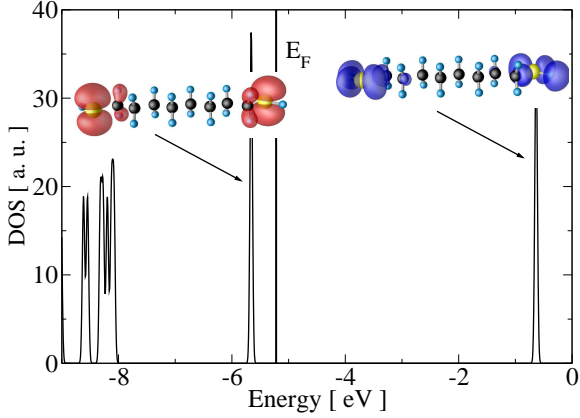}
\end{center}
\caption{[8]-alkane (octane) molecule:
DOS and charge density isosurface plots for the relevant molecular states of
the isolated molecule. $E_\mathrm{F}$ denotes the position of the Fermi level
for the isolated molecule.}
\label{fig7}
\end{figure}

\begin{figure}[htb]
\begin{center}
\includegraphics[width=5.5cm,clip=true,angle=-90]{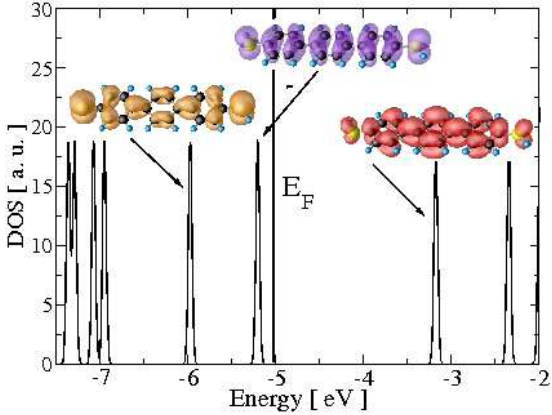}
\end{center}
\caption{1,4-[3]-phenyl (tricene) molecule:
DOS and charge density isosurface plots for the relevant molecular states of
the isolated molecule. $E_\mathrm{F}$ denotes the position of the Fermi level
for the isolated molecule.}
\label{fig8}
\end{figure}

As an example of different transport regimes, we consider spin-valves made from Ni leads 
and a molecular spacer which is either [8]-alkane-dithiolate (octane-dithiolate) or
1,4-[3]-phenyl-dithiolate (tricene-dithiolate). 
A schematic DOS and the charge density isosurfaces of the HOMO and LUMO 
states for the isolated molecules are presented in figures \ref{fig7} and \ref{fig8}. 

The two molecules present rather different characteristics. The HOMO-LUMO gap \cite{gap}
is about 2.5~eV for tricene and almost double (5~eV) for octance. In addition while in
tricene the charge density of the first two HOMO levels and the LUMO is extended over the
whole molecule, in octane this is predominantly localized around
the S atoms of the thiol groups. We then expect that the octane and the tricene 
will form respectively TMR and GMR devices, as actually found in our calculations \cite{Smeagol1}. 

\begin{figure}[htb]
\begin{center}
\includegraphics[width=6.5cm,clip=true]{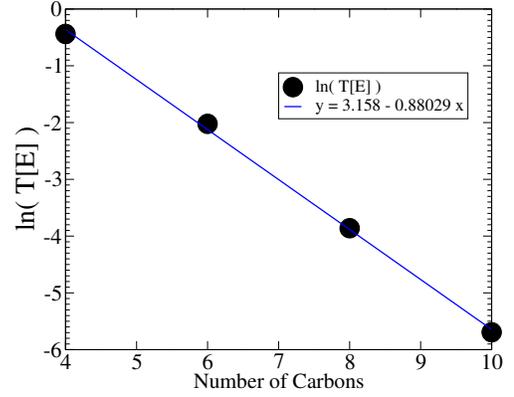}
\end{center}
\caption{$\ln[T(E_\mathrm{F})]$ as a function of the number of C atom for
[n]-alkane-dithiolate attached to Ni leads. The black circles correspond to our 
calculated values, and the solid line is our best linear fit. Here the spin-valve is 
in the parallel configuration.}
\label{fig9}
\end{figure}

Let us consider octane first. The zero-bias transmission coefficient of the Ni/octane/Ni junction 
presents a sharp peak at $E_\mathrm{F}$ that scales exponentially with the number of alkane groups
$T\propto\mathrm{e}^{-\beta n}$ ($\beta\sim0.88$). This is demonstrated in figure \ref{fig9}
for the parallel configuration and confirms that device is in a tunneling regime. Note that the method
is accurate down to a conductance of $\sim50$pS.
Interestingly the exponent is similar to that found for the same molecule attached to 
gold (111) surfaces \cite{Sankey}. The coupling between the thiol groups and the electrodes 
is strong and it gives rise to a small spin-polarization of the two S atoms. However since
both the HOMO and LUMO states are strongly localized at the thiol groups, at the Fermi level 
there is no molecular state extending through the entire structure. Hence the 
Ni/octane/Ni junction presents the features of a TMR spin-valve.

\begin{figure}[htb]
\begin{center}
\includegraphics[width=7.5cm,clip=true]{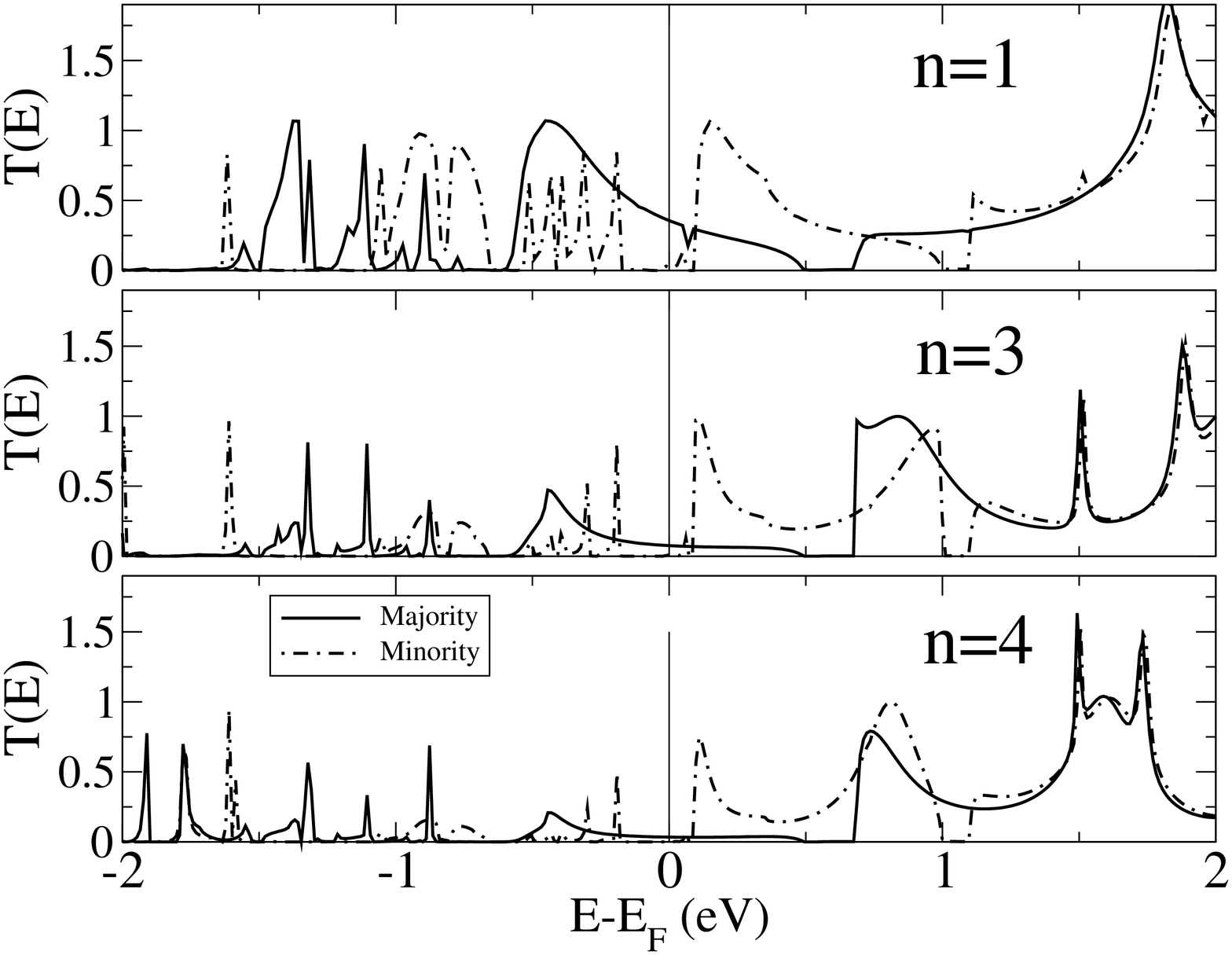}
\end{center}
\caption{Transmission coefficient as a function of energy for 1,4-[n]-phenyl-dithiolate
molecules with n=1, 3, and 4. Here the spin-valve is in the parallel configuration. A similar
scaling of the transmission coefficient as a function of n is found for the antiparallel 
configuration.}
\label{fig10}
\end{figure}

The case of 1,4-[n]-phenyl-dithiolate is different, in particular we find that the current does 
not scale sensibly with the number of phenyl groups in the molecule. In figure \ref{fig10} 
we present the transmission coefficient as a function of energy of 1,4-[n]-phenyl-dithiolate 
for 1, 3 and 4 phenyl rings when the magnetization vectors of the leads are parallel to each other.
It is clear that, although there is a reduction of the transmission coefficient as a function
of the number of rings, this remains close to unity for most of the energy range, and certainly 
there is not an exponential decay. Therefore in this class of molecules the transport seems 
to be appropriately described by a coherent resonant tunneling mechanism through
extended molecular states. 

This picture is enforced by the fact that the
zero-bias transmission coefficient approaches unity for energies close to the leads Fermi 
level. In addition, from the study of the evolution of the orbital resolved density of states 
as a function of the distance between the thiol group and the electrodes \cite{Smeagol1} 
we identify such a resonant state as the HOMO state of tricene. However it is worth 
mentioning that this appears rather broad and spin-split, because of the strong coupling 
with the $d$ orbitals of the leads. In conclusion all this suggests that a Ni/tricene/Ni spin-valve 
behaves as metallic spin-valve.

\begin{figure}[htb]
\begin{center}
\includegraphics[width=7.5cm,clip=true]{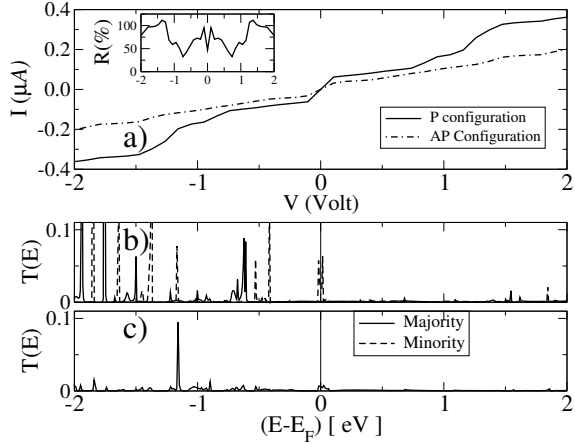}
\end{center}
\caption{a) $I$-$V$ characteristic, and zero bias transmission coefficients
for the b) parallel and c) antiparallel configuration of an octane-based
Ni spin-valve. In the antiparallel case the transmission coefficient is identical
for both the spin directions. In the inset we present the corresponding MR ratio.
$E_\mathrm{F}$ is the position of the Fermi level of the Ni leads. From reference 
\cite{Smeagol2}.}
\label{fig11}
\end{figure}

\begin{figure}[htb]
\begin{center}
\includegraphics[width=7.5cm,clip=true]{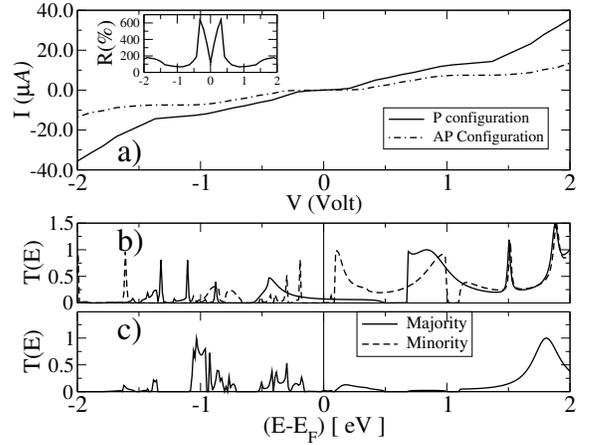}
\end{center}
\caption{a) $I$-$V$ characteristic, and zero bias transmission coefficients
for the b) parallel and c) antiparallel configuration of a tricene-based
Ni spin-valve. In the antiparallel case the transmission coefficient is identical
for both the spin directions. In the inset we present the corresponding MR ratio.
$E_\mathrm{F}$ is the position of the Fermi level of the Ni leads. From reference 
\cite{Smeagol2}.}
\label{fig12}
\end{figure}

The $I$-$V$ characteristics of both the molecules are strongly non-linear with the bias
and consequently also the GMR ratio suffers this non-linearity. In figures \ref{fig11} 
and \ref{fig12} we present the $I$-$V$ curves, the GMR ratio and the zero-bias transmission
coefficient for the two molecules. 

Again consider octane first. Here the transmission coefficient at zero bias is dominated by
a number of sharp peaks in the parallel configuration, which get considerably suppressed
in the antiparallel one. In particular a minority peak appears at the Fermi level and it is
mostly responsible for the low-bias conductivity. Such sharp peaks in the transmission coefficient
are usually a signature of resonant states at the interface between the leads and the insulating
media. At resonance they can carry a considerable current, however small bias and disorder 
are usually rather effective in suppressing their contribution to the current. Evidence 
for these surface states has been already provided for conventional magnetic tunneling junctions
both experimentally \cite{Inter-Exp} and theoretically \cite{Tsymbal}. 

The $I$-$V$ characteristic is rather linear for the antiparallel configuration but presents 
a non-trivial slope for the parallel case. This gives rise to a non-monotonic dependance
of the GMR ratio over the bias. Importantly both the layer resistance and the TMR ratio are 
in the same range as in recent experiments on octane-based Ni spin-valves \cite{Ralph}. 
However, a direct comparison with experiments is difficult since the actual number of
molecules bridging the two electrodes is not known with precision. Moreover a degradation of the
GMR signal due to spin-flip and electron-phonon scattering, misalignment of the magnetization of the
contacts and current shortcut through highly conductive pin-holes, can drastically reduce 
$R_\mathrm{GMR}$ in actual samples.

In contrast in the case of metallic (tricene) junctions the transmission coefficient at zero bias presents
a much higher transmission and is a rather smooth function of the energy. In the parallel case most
of the transmission at $E_\mathrm{F}$ is due to the majority spin, while the contribution of the 
minority is significant only for energies at about 200~meV above $E_\mathrm{F}$. In the antiparallel 
configuration such transmission at the Fermi level is strongly suppressed and the resulting
current is considerably lower. This produces an extremely large GMR ratio for small bias, 
exceeding 600\%. 

\begin{figure}[htb]
\begin{center}
\includegraphics[width=7.5cm,clip=true]{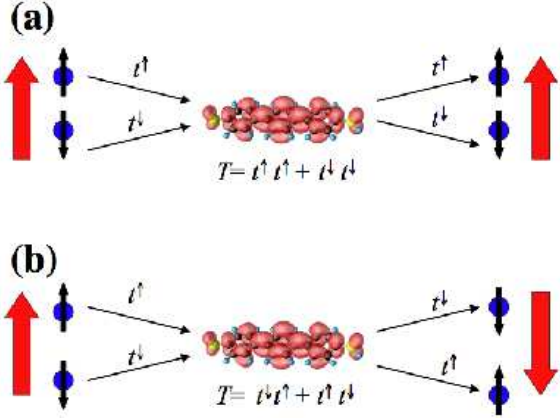}
\end{center}
\caption{Scheme of the spin-transport mechanism through a single molecular state. 
$t^\uparrow(E)$ ($t^\downarrow(E)$) is the majority (minority) spin hopping integral 
from one of the leads to the molecular state. Neglecting quantum interference, 
in the parallel case (a) the total transmission coefficient is simply 
$T=(t^{\uparrow})^2+(t^{\downarrow})^2$, while in the antiparallel (b) 
$T=2t^{\uparrow}t^{\downarrow}$. Note that if either $t^\uparrow$ or
$t^\downarrow$ vanishes, the current in the antiparallel configuration
will also vanish (infinite GMR).}
\label{fig13}
\end{figure}

An interesting feature is that, in first approximation, the transmission coefficient at zero bias for the 
antiparallel state appears to be a convolution of those for the majority and minority spin in the parallel case. 
This finding can be qualitatively understood in terms of transport through a single molecular state 
(see figure \ref{fig13}). Let $t^\uparrow(E)$ be the majority spin hopping integral from one
of the leads to the molecular state, and $t^\downarrow(E)$ the
same quantity for the minority spins. Then, neglecting multiple
scattering (i.e. all interference effects), the total transmission coefficients
of the entire spin-valve in the parallel state can be written
$T^{\uparrow\uparrow}(E)=(t^{\uparrow})^2$ and
$T^{\downarrow\downarrow}(E)=(t^{\downarrow})^2$ respectively for
the majority and minority spins. Similarly the transmission in the
anti-parallel configuration is
$T^{{\uparrow\downarrow}}(E)=T^{\downarrow\uparrow}(E)=
t^{\uparrow}t^{\downarrow}$. Thus $T^{{\uparrow\downarrow}}(E)$ is
 a convolution of the transmission coefficients for
the parallel case $T^{\uparrow\downarrow}\propto
\sqrt{{T^{\uparrow\uparrow}}T^{\downarrow\downarrow}}$.

\begin{figure}[htb]
\begin{center}
\includegraphics[width=7.5cm,clip=true]{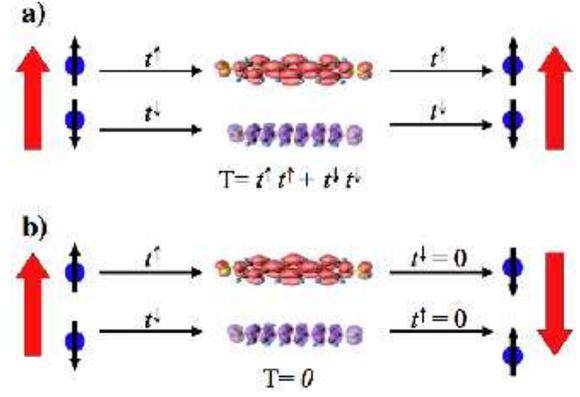}
\end{center}
\caption{Scheme of the spin-transport mechanism through two energetically closely spaced 
molecular states. The first state (red) couples only to the majority spin band,
while the minority spin couple only to the purple state. In the parallel case
one finds $T^{\uparrow\uparrow}(E)\ne0$ and $T^{\downarrow\downarrow}(E)\ne0$ 
but in the antiparallel $T^{\uparrow\downarrow}(E)=0$. We then expect an infinite
GMR for such an energy window.}
\label{fig14}
\end{figure}

An extreme case is when only one spin couples to the molecular
state. Then the total transmission in the antiparallel case is
identically zero since either $t^{\uparrow}$ or $t^{\downarrow}$
vanishes. This is the most desirable situation in real devices
since, in principle, an infinite $R_\mathrm{GMR}$ can be obtained.
Note that in this situation the system leads+molecule behaves as
a half-metal although the two materials forming the device are not
half-metals themselves. An even more extreme situation is when for a particular energy window 
the transport is through two distinct molecular states, which are respectively
coupled to the majority and minority spin only. This may happen for instance 
due a particular symmetry of the molecular anchoring groups. Then in this
energy window one will find $T^{\uparrow\uparrow}(E)\ne0$,
$T^{\downarrow\downarrow}(E)\ne0$ but
$T^{\uparrow\downarrow}(E)=0$ (see figure \ref{fig14}).

\begin{figure}[h]
\begin{center}
\includegraphics[width=7.5cm,clip=true]{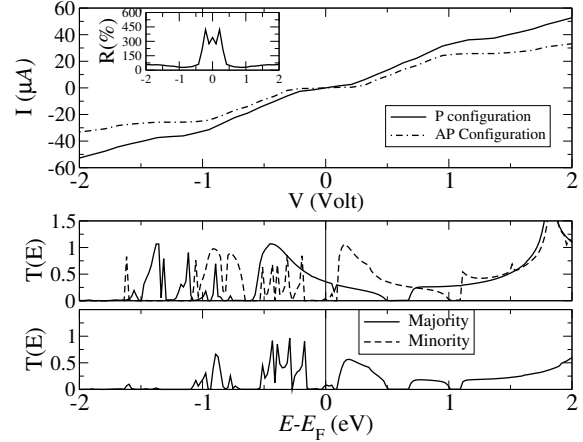}
\end{center}
\caption{Transport properties for a 1,4-phenyl molecule attached to Ni (100) surfaces
through a S group. The top panel shows the $I$-$V$ characteristics for both the parallel and
antiparallel alignment of the leads and the inset the corresponding GMR ratio. The lower
panel is the transmission coefficient at zero bias as a function of energy. Because of
spin-symmetry, in the antiparallel case we plot only the majority spin. Reprinted with permission from
\cite{Smeagol1} A.R. Rocha et al., Phys. Rev. B {\bf 73},  085414 (2006). Copyright American Physical
Society 2006.}
\label{fig15}
\end{figure}

\begin{figure}[h]
\begin{center}
\includegraphics[width=7.5cm,clip=true]{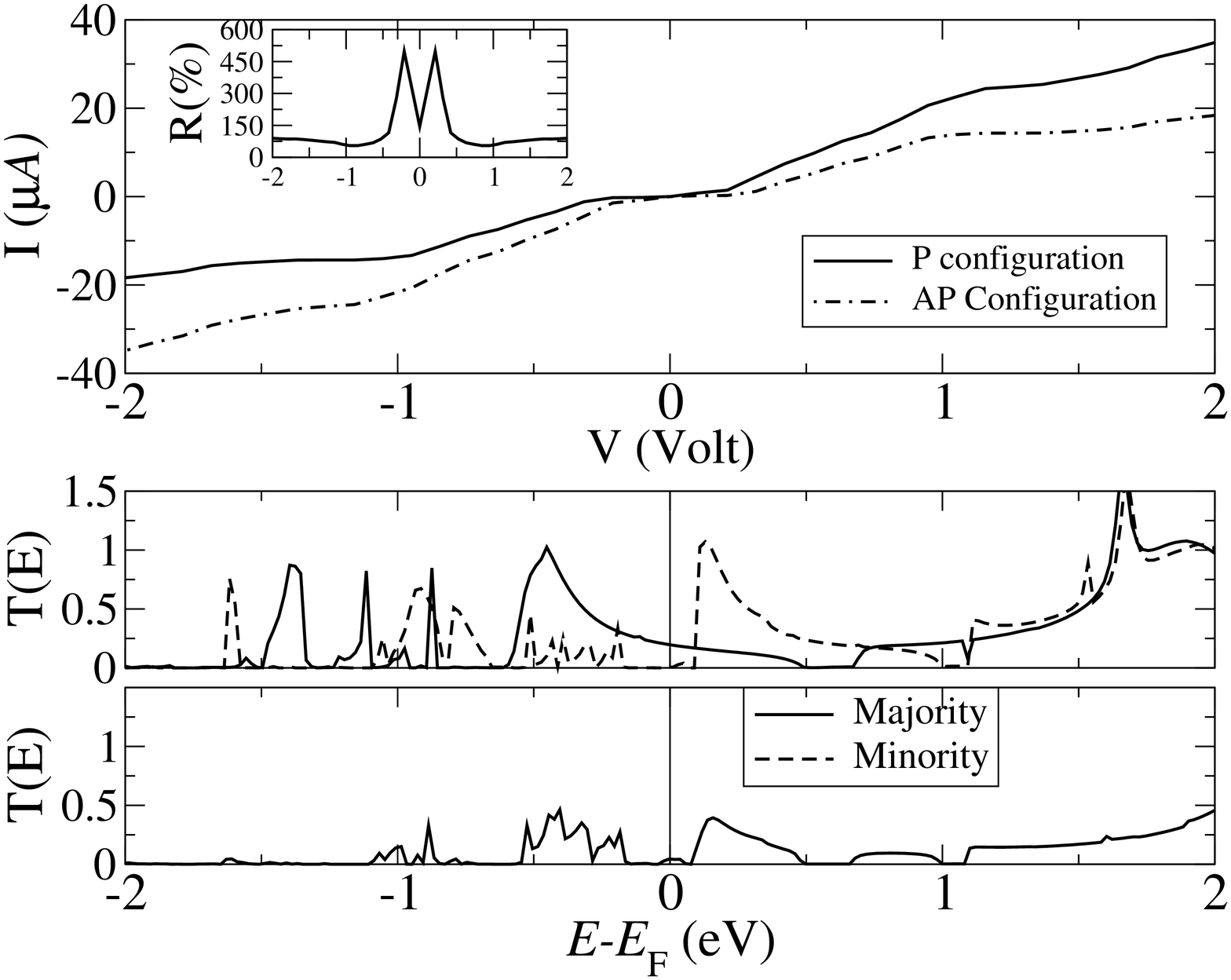}
\end{center}
\caption{Transport properties for a 1,4-phenyl molecule attached to Ni (100) surfaces
through a Se group. The top panel shows the $I$-$V$ characteristics for both the parallel and
antiparallel alignment of the leads and the inset the corresponding GMR ratio. The lower
panel is the transmission coefficient at zero bias as a function of energy. Because of
spin-symmetry, in the antiparallel case we plot only the majority spin. Reprinted with permission from
\cite{Smeagol1} A.R. Rocha et al., Phys. Rev. B {\bf 73},  085414 (2006). Copyright American Physical
Society 2006.}
\label{fig16}
\end{figure}

\begin{figure}[h]
\begin{center}
\includegraphics[width=7.5cm,clip=true]{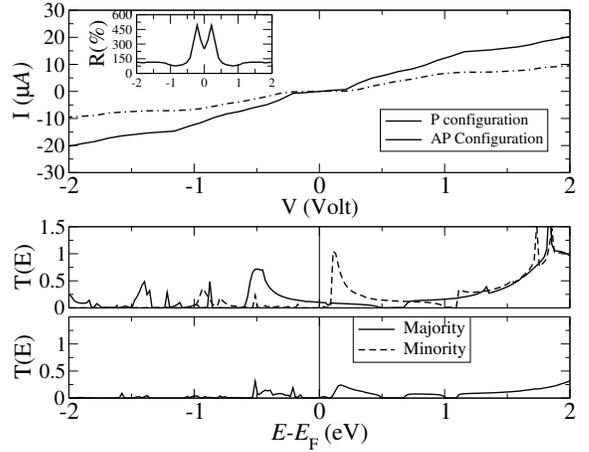}
\end{center}
\caption{Transport properties for a 1,4-phenyl molecule attached to Ni (100) surfaces
through a Te group. The top panel shows the $I$-$V$ characteristics for both the parallel and
antiparallel alignment of the leads and the inset the corresponding GMR ratio. The lower
panel is the transmission coefficient at zero bias as a function of energy. Because of
spin-symmetry, in the antiparallel case we plot only the majority spin. Reprinted with permission from
\cite{Smeagol1} A.R. Rocha et al., Phys. Rev. B {\bf 73},  085414 (2006). Copyright American Physical
Society 2006.}
\label{fig17}
\end{figure}

The fact that in good approximation $T^{\uparrow\downarrow}\propto
\sqrt{{T^{\uparrow\uparrow}}T^{\downarrow\downarrow}}$ can be used for enhancing the GMR
ratio. Consider for instance the case when $T^{\uparrow\uparrow}\gg T^{\downarrow\downarrow}$. 
Then we have $T_\mathrm{P}\approx T^{\uparrow\uparrow}$ and 
$T_\mathrm{AP}\approx \sqrt{{T^{\uparrow\uparrow}}T^{\downarrow\downarrow}}$
for the total transmission coefficients respectively of the parallel and antiparallel 
configurations. Clearly a reduction of $T^{\downarrow\downarrow}$ will
produce a considerable reduction in the transmission of the antiparallel alignment, leaving
almost unchanged that of the parallel one. We have explored this avenue \cite{Smeagol1} by
replacing the thiol group with either a Se or a Te atoms. These provide a rather strong
bond, although generally the bond length is increased due to the larger atomic radius of the 
anchoring atom. 
The transmission, $I$-$V$ characteristics and GMR for 1,4-benzene anchored to Ni
via S, Se or Te are presented in figures \ref{fig15}, \ref{fig16} and \ref{fig17}.
Clearly the pictures show a rather dramatic increase of the GMR demonstrating
that the GMR signal can be tuned by an appropriate choice of the anchoring chemistry.

Finally it is worth mentioning that a severe bias dependance of the GMR ratio, with also the possibility of
negative values, has been recently predicted with either semi-empirical \cite{Dalgleish} and
{\it ab initio}  \cite{GuoNi} methods.

\subsection{Carbon Nanotubes}

Carbon nanotubes are almost defect-free graphene sheets rolled up to form one-dimensional
molecules with enormous aspect ratios \cite{Dress}. Their conducting state (metalicity) depends on
their chirality, however in the metallic configuration they are ideal conductors with a remarkably long
phase-coherence length \cite{Heer,todorovCN}. An important aspect is that the relevant physics at
the Fermi level is entirely dominated by the $p_z$ orbitals, which are radially aligned with respect 
to the tube axis. These include the bonding properties with other materials and between tubes. 
Therefore carbon nanotubes appear as an ideal playground for investigating both GMR and TMR
through molecules. In fact one can expect that two tubes with different chirality will bond to a magnetic
surface in a similar way, allowing us to isolate the effects of the molecule from that of the contacts.
Indeed TMR-like transport through carbon nanotubes has been experimentally reported by several 
groups \cite{bruce,CNT1,CNT2,CNT4,CNT5,CNT6,CNT7,CNT8}.

\begin{figure}[ht]
\begin{center}
\includegraphics[width=2.2cm,clip=true,angle=-90.0]{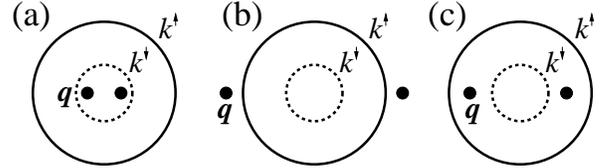}
\end{center}
\caption{Fermi surfaces of an armchair carbon nanotube and of a magnetic
transition metal. The Fermi surface of the nanotube consists in two points
$k_{\mathrm F}^{\mathrm N}=q$,
symmetric with respect to the $\Gamma$ point. The Fermi surface of a
transition magnetic metal consists of two spheres (for $\uparrow$ and $\downarrow$ spins) whose
different diameters depend on the exchange field. The three possible scenarios
discussed in the text: 
(a) $q < k_{\mathrm F}^\downarrow < k_{\mathrm F}^\uparrow$,
(b) $k_{\mathrm F}^\downarrow < k_{\mathrm F}^\uparrow < q$,
(c) $k_{\mathrm F}^\downarrow < q < k_{\mathrm F}^\uparrow$.}
\label{fig18}
\end{figure}

Why would one expect a large GMR from a carbon nanotube?
To answer this question we use an argument derived by Tersoff \cite{Tersoff} and then 
subsequently refined \cite{MaxJTersoff} for explaining the contact resistance between a C-nanotube
and an ordinary metal. Consider for simplicity an armchair nanotube (metallic).
The Fermi surface of such tube consists only of two points, symmetric with respect to
$\Gamma$ in the 1D Brillouin zone (see figure \ref{fig18}). The Fermi wave-vector is then
$k_{\mathrm F}=2\pi/3z_0$ with $z_0=d_0\sqrt{3}/2$ and $d_0$ the C-C bond 
distance ($d_0$=1.42~\AA). 

Assume now for simplicity that our magnetic metal is an exchange-split free-electron gas,
whose band-energy is
\begin{equation}
E^\sigma_k=\frac{\hbar^2k^2}{2m}+\sigma \Delta/2
\;{,}
\end{equation}
with $\sigma=-1$ ($\sigma=+1$) for majority (minority) spins and $\Delta$ the 
exchange energy. The spin-dependent Fermi wave-vectors are then respectively
$k_{\mathrm F}^\uparrow=\sqrt{2m(E_{\mathrm F}+\Delta/2)}/\hbar$ and
$k_{\mathrm F}^\downarrow=\sqrt{2m(E_{\mathrm F}-\Delta/2)}/\hbar$.

The transport through an interface between such a magnetic metal and the
nanotube is determined by the overlap between the corresponding Fermi surfaces.
Three scenarios are possible. First the Fermi-wave vector of the carbon nanotube is smaller than both 
$k_{\mathrm F}^\uparrow$ and $k_{\mathrm F}^\downarrow$
(see figure \ref{fig18}a). In this case in the magnetic metal there is always a $k$-vector that matches the 
Fermi-wave vector of the nanotube for both spins. Therefore both spins can be
injected into the tube and the total resistance will be small and
spin-independent.

Secondly the Fermi-wave vector of the carbon nanotube is larger than both 
$k_{\mathrm F}^\uparrow$ and $k_{\mathrm F}^\downarrow$
(see figure \ref{fig18}b). Now no states are available in the metallic contacts whose wave-vectors match
the Fermi wave-vector of the carbon nanotube. For zero-bias and zero-temperature
the resistance is extremely large. Nevertheless upon increasing the temperature,  phonon assisted transport 
becomes possible. Spin electrons can be scattered out of the Fermi surface into states with large 
longitudinal momentum. At temperature $T$ the fraction of electrons with energy 
above $E_{\mathrm F}$ is simply proportional to the Fermi distribution function. 
However, because of the exchange energy, spin-up electrons will
possess higher momentum than spin-down. Therefore one can find more
spin-up states with a longitudinal momentum matching the one of the nanotube than 
spin-down states. This  potentially gives a temperature-induced spin-dependent resistance. 

Finally if the Fermi wave-vector of the carbon nanotube is larger than
$k_{\mathrm F}^\downarrow$ but smaller than $k_{\mathrm F}^\uparrow$
(see figure \ref{fig18}c), only the majority electrons
can enter the nanotube. The system becomes fully spin-polarized and the Fermi surface
matching replaces the bonding spin-selectivity encountered in the previous section.
Also in this case a spin-valve structure made by magnetic contacts and carbon nanotube as
spacer is predicted to show infinite GMR at zero temperature. An increase of the temperature will
produce a degradation of the polarization since minority spins can be thermally scattered outside
their Fermi surface, and therefore contribute to the transport.

Two important aspects must be pointed out. First all these considerations are
based on the assumption of perfectly crystalline systems. This may not be true
in reality and the effects of breaking the translational invariance must be
considered. From a qualitative point of view disorder smears the Fermi
surface and eventually may produce some states with large longitudinal momentum.
This will improve the conductance through the nanotube, even if its
spin-polarization will be in general dependent on the nature of disorder. 

Secondly, our heuristic argument does not necessarily apply to the case of transition metal
contacts, where the spin-selectivity arising from the different bonding nature of the two spin-subands can
play an important r\^ole. Clearly more realistic bandstructure calculations are needed. 
These however are rather problematic. In addition to the need of simulating transition metal
leads one has to consider rather large supercell for a reasonable description of the 
nanotube/metal interface.

For this reason most of the calculations to date have used simple tight-binding 
models without self-consistent procedures \cite{Mehrez,Kromp1,Kromp2}. These roughly 
agree on the possibility of large GMR ratios in transition metals contacted nanotubes,
although the actual values predicted are somehow affected by the different methods and the
contact geometry. 

\subsection{Long molecules and phonons}
As mentioned previously electronic transport in polymers and long molecules offer a 
considerable higher level of complexity when compared to small ballistic molecules. The first consideration
is that the transport is often driven by strong electron-phonon interaction, therefore the relevant
current carriers are some correlated electronic-vibronic states (polarons, bi-polarons, etc.). Moreover
in the case of spin-transport one has to take into account the likely presence of paramagnetic centers
\cite{EPR}. They usually appear in low concentrations, but they become relevant to the spin-dynamics
in long molecules. These two features add to the large scale of the system and make the problem 
of spin-transport in large organic molecules intractable with {\it ab initio} methods. This is the main reason
why to date only simple Hamiltonian models have been used.

When the electron-phonon interaction is weak, or alternatively the molecule is rather short
in such a way that polaronic-type of transport is not dominant, then phonon absorption/emission has only 
the effect of smearing the transmission coefficient. In spin-valves made from transition metals such a 
smearing is likely to result in a reduction of the spin-polarization of the device and therefore of the
GMR \cite{jean}. However quantitative predictions are difficult, since also in this case a detailed description of the 
electronic structure of the electrodes and of the bonding with the molecule is essential. For instance 
calculations of spin-injection into short strands of DNA obtained with simple single-orbital tight-binding models 
and some parameterization of the metal/molecule bonding, report both an enhancement \cite{Max} and a 
reduction \cite{wang} of the GMR with the bias.

When the electron-phonon interaction is strong and the molecules are rather long, then the ground
state of the molecule is some correlated electronic-vibronic state. In this case the situation is
more complex and to the best of our knowledge no transport calculations have been carried out
to date. Some interesting insights come from the investigation of the ground state of polymers in 
contact with magnetic metals. Xie and co-workers  \cite{Bishop} investigated a non-degenerate 
polymer in contact with a model metal reproducing either a magnetic transition metal or
manganite. Interestingly they found that when no charge is transferred from the metal to the polymer,
this remains metallic with a rather uniform distribution of the charge across the interface. 
In contrast charge transfer promotes the formation of spinless bipolarons in the polymers. An analysis of
the DOS of the whole structure further suggest that the bipolaron formation drastically reduces the
spin polarization of the whole system.

\section{More exotic phenomena}

\subsection{Molecular Magnets}

Extremely interesting features appear in quantum transport through molecules when
the conducting electrons interact with some internal molecular degrees of freedom. This is for instance
the case of vibrational levels, where electron-phonon interaction manifests itself with sharp changes 
in the slope of the $I$-$V$ curve, steps in $\mathrm{d}I/\mathrm{d}V$ and in peaks in
$\mathrm{d}^2I/\mathrm{d}V^2$.
It is therefore natural to speculate about similar effects associated to internal magnetic degrees 
of freedom, i.e. the molecular spin \cite{kimkim}. Until recently the experimental activity
was focussed on reproducing at the molecular level effects already demonstrated in quantum dots,
such as the Zeeman splitting of the Coulomb-blockade as well as the Kondo effect \cite{kondo1,kondo2}.
More recently, advances in the chemical functionalization of magnetic molecules have allowed
the formation of stable bonding between the molecules and metallic surfaces \cite{cornia,mannini}, 
thus the construction of two and three terminal devices \cite{mmtrans1,mmtrans2}.

Magnetic molecules \cite{robertareview} are molecules comprising a number 
of transition metal ions magnetically 
coupled to each other in such a way to give rise to a global net spin $S$. They are usually
described by the following spin Hamiltonian $H_0$
\begin{equation}
H_0=DS_z^2+g\mu_\mathrm{B}H_zS_z\;,
\label{0field}
\end{equation}
where $D$ ($D<0$) is the zero field splitting constant and $H_z$ is the $z$ component of an 
external magnetic field ($\mu_\mathrm{B}$ is the Bohr magneton). The term proportional to
$S_z^2$ lift the degeneracy of the spin-multiplet and the energy levels can be labeled by
the magnetic quantum number $M_S$, with $-S\le M_S \le S$. 

If the dynamics is solely determined
by the Hamiltonian $H_0$ then a molecule prepared in a given $M_S$ state will remain in such 
state. However a perturbation $H_1$, which does not commute with $H_0$ will create
mixing between the $M_S$ states, thus transition between levels with different $M_S$ will
be possible. Selection rules for these transitions are given by the particular symmetry of the perturbation,
and for instance transverse anisotropy 
\begin{equation}
H_1=E(S_x^2-S_y^2)\;,
\label{trananiso}
\end{equation}
allows transitions $M_S\rightarrow M_S\pm 2n$ with $n$ an integer. Evidence of quantum tunneling of
the magnetization for magnetic molecules are now numerous \cite{robertareview}.

Let us now consider the case of electronic transport through such a molecule and in particular the
case of weak coupling between the molecule and the current/voltage electrodes, which is the most
likely situation in actual experiments \cite{mmtrans1,mmtrans2}. In general the ground state for
a charged molecule will be different from that of its neutral state, and so will be the excitation spectrum.
This means that sequential tunneling can potentially leave the molecule in some excited state
not allowed to relax by the selection rules. Importantly also the opposite is true, namely that a charged
exited molecule can relax to a state which does not allow electron transfer to the
electrodes because of the selection rules. This may lead to a complete suppression of the current \cite{mmtrans1}.

\begin{figure}[h]
\begin{center}
\includegraphics[width=8.5cm,clip=true,angle=0.0]{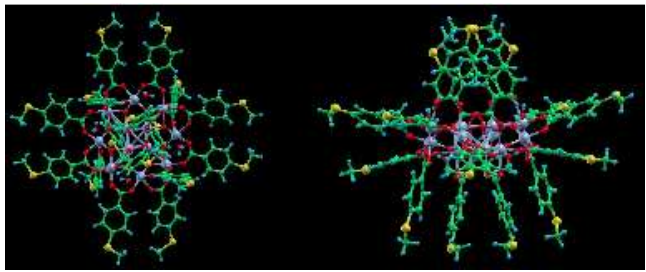}
\end{center}
\caption{Top and side views of a ball-and-stick model for a Mn$_{12}$ magnetic molecule. In this case we consider 
[Mn$_{12}$O$_{12}$(O$_2$C-R-SAc)$_{16}$(H$_2$O)$_{4}$] where R=$\{$
C$_6$H$_4$, C$_{15}$H$_{30}$ $\}$. Colour code: green=carbon, yellow=sulphur, blue=hydrogen, red=oxygen and violet=manganese.}
\label{fig19}
\end{figure}

Two experimental works have been published so far on transport through magnetic molecules
\cite{mmtrans1,mmtrans2}. They both consist in a transistor geometry where a single Mn$_{12}$ molecule is trapped. Mn$_{12}$, [Mn$_{12}$O$_{12}$(CH$_3$COO)$_{16}$(H$_2$O)$_4$],  is perhaps the prototype of all molecular magnets. The 12 Mn ions occupy three inequivalent atomic sites, namely two Mn$^{3+}$
and one Mn$^{4+}$. The Mn ions with different valence couple antiferromagnetically to each other resulting
in a total $S=10$ ground state. In one case the molecule has been functionalized (see figure \ref{fig19}) with thiol groups in order to form stable contacts with the gold surfaces of the electrodes. The crucial aspect of both these experiments is
the discovery of fingerprints of the molecular state into the $I$-$V$ characteristics of the device. In particular
negative differential conductance, complete current suppression \cite{mmtrans1} and non-trivial dependence
of the peaks in the differential conductance over an applied magnetic field \cite{mmtrans2}, can all be 
interpreted as a result of the internal spin-dynamics of the molecule. 

Several model calculations have been performed for explaining the various features of these 
experiments \cite{weg1} and proposing a new setup where one of the electrodes is a ferromagnet
\cite{tim1,tim2}. These are based on the spin-Hamiltonian of equations (\ref{0field}) and (\ref{trananiso})
and an additional term that takes into account the coupling with the electrodes. Then the transport 
calculation is performed within the standard master equation formalism familiar to transport through
interacting quantum dots \cite{master}. Although these calculations are certainly important since they
demonstrate the mechanisms behind the various effects, they depend heavily on the specific choice
of the parameters used. For this reason {\it ab initio} simulations would be very important. For instance
questions about the actual charging state of the molecule under bias, the strength of the interaction
between the conducting electrons and the Mn ions, and the electrical response of the whole system
are likely to find an answer with DFT calculations. The problem is indeed complicated and even the
simple evaluation of the ground state of magnetic molecules from DFT is not trivial \cite{ped1,ped2}. 
For this reason no first principle calculations of transport through magnetic molecules have been performed 
to date. However we believe that this is an extremely challenging field where order-N capability, strong correlation
and scalable quantum transport schemes can find a common playground.

Finally we want to discuss the possibility of constructing all-molecular spin-valves, i.e. 
spin-valves where both the spin-injector and spin-detector are molecules or part of the same
molecule and  the metallic current/voltage electrodes have the only function of electron reservoirs. 
Also in this case the expectation is to detect electrically the magnetic state of the molecule, however
for spin-valve operation we also demand the possibility of switching reversibly between two spin-configurations by applying a magnetic field. An interesting proposal is that of using dicobaltocene
molecules \cite{2Co}. These belong to the metallocene family and are characterized by two Co ions separated by a spacer that can be chemically engineered. 

DFT calculations \cite{2Co} show that dicobaltocene attached to gold electrodes in its ground state
is stabilized by superexchange in an antiferromagnetic configuration, i.e. the local moments of the two
Co ions are antiparallel to each other. As in a conventional spin-valve a parallel alignment can
be obtained by applying a large magnetic field (around 20~T for a C$_2$H$_4$ spacer), and the calculation
shows a rather large GMR ratio. This is suggestive of the possibility of all-molecular spin-valves and future 
experiments in this area are certainly welcome.

\subsection{d$^0$ ferromagnetism and magnetic proximity}

We wish to close this review with a brief discussion of two recently discovered intriguing 
phenomena, that challenge our current understanding of ferromagnetism and may offer a new
playground for spin-transport. These are d$^0$ ferromagnetism and magnetic proximity effect. 
The measurement of long-range ferromagnetic order in materials not containing ions with either
$d$ or $f$ electrons, therefore without an obvious way to produce a local magnetic moment, is the
common factor of these two aspects of magnetism.

Let us consider first the case of d$^0$ ferromagnetism \cite{d0ferro}. This is the intrinsic 
ferromagnetism of usually highly defective materials, which do not contain any partially filled $d$ 
or $f$ shells. Among the several examples we wish to mention irradiated graphite and
fullerenes \cite{makarova}, nonstoichiometric CaB$_6$ \cite{CaB6}, and HfO$_2$ thin films \cite{HfO2}.
A common explanation for the magnetism in all these materials is not available at present. Note that
one has to explain both the formation of a magnetic moment and the long-range coupling between 
moments. Generally the moments are associated to intrinsic defects. Paramagnetic defects in organic
materials are not uncommon \cite{EPR} and strongly correlated molecular orbitals associated to
vacancies have been suggested for magnetic oxides \cite{Das,CaO1}. However the demonstration of
long range coupling remains elusive. For instance recent DFT calculations \cite{CaO2} demonstrate, 
at least for CaO, that the defect concentration needed for long range ferromagnetism is three orders of 
magnitude larger than that obtainable at equilibrium. Finally we wish to mention that ferromagnetism 
originating from $p$-shells has been suggested for oxygen mixed valence compounds such as 
Rb$_4$O$_6$ \cite{RbO}.

In contrast to d$^0$ ferromagnetism magnetic proximity is not an intrinsic material property and
it does require the presence of a magnetic material. The basic idea is 
quite simple: there is always some charge transfer at the contact between a conducting 
molecule and a metal associated with the alignment of their respective chemical potentials. 
In a ferromagnet some degree of spin transfer accompanies the charge 
transfer giving rise to an induced magnetic moment. Note that the magnetic proximity effect does 
not imply intrinsic ferromagnetism and no spin aligning potentÄial exists in the non-magnetic material. 
This means that a magnetic moment is detected only when a second ferromagnetic material is present 
and when good contact is made. 

A first indirect evidence of this effect was found in explaining the unaccounted magnetization
in a carbon meteorite rich of magnetic inclusions \cite{meteo}. Then Ferreira and Sanvito
derived a close system of equations for the induced magnetic moment and for the energetic
of a carbon nanotube deposited on a magnetic surface \cite{ss_ferreira}. The calculation, based
on a simple tight-binding model, revealed that induced magnetic moments 
of the order of 0.1~$\mu_\mathrm{B}$ per carbon atom in contact can be achieved
at room temperature. This paved the way for a more controlled set of experiments.

\begin{figure}[ht]
\begin{center}
\includegraphics[width=7.5cm,clip=true,angle=0.0]{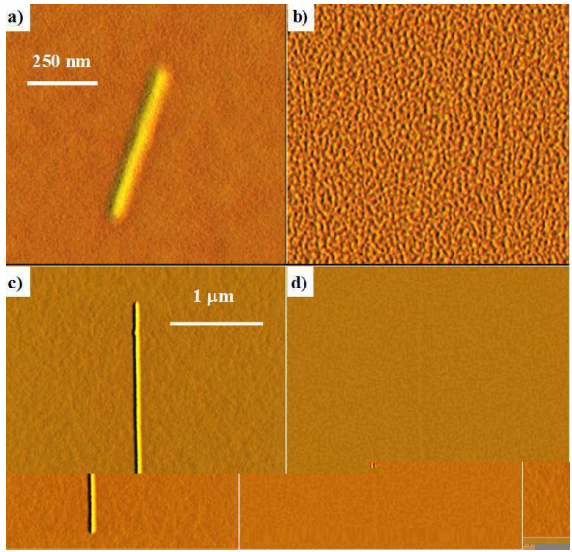}
\end{center}
\caption{AFM images of a carbon nanotube on copper a) and silicon c) substrate and their
corresponding MFM scans b) and d). The MFM images do not show any magnetic contrast
indicating no induced magnetic moment.}
\label{fig22}
\end{figure}

\begin{figure}[ht]
\begin{center}
\includegraphics[width=7.5cm,clip=true,angle=0.0]{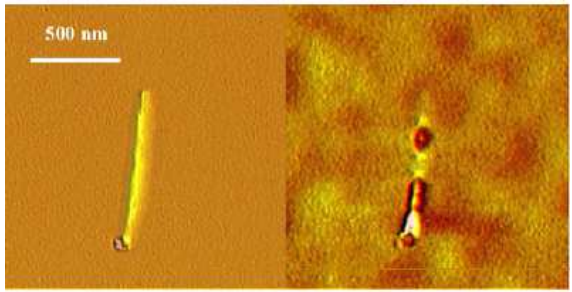}
\end{center}
\caption{AFM images of a carbon nanotube on cobalt substrate (left) and the
corresponding MFM scans (right). The MFM image shows magnetic contrast
indicating an induced magnetic moment.}
\label{fig23}
\end{figure}

Experimentally the problem is to detect the small induced moment over a huge background 
coming from the ferromagnetic substrate. One possible strategy is that of measuring the
stray field around the nanotube once this is placed on a smooth thin film.
A uniformly magnetized thin film creates no stray field whatever its direction of magnetization. 
In contrast a magnetized nanotube will produce a stray field, which will be directly detectable.
This idea was exploited in the experiments from C\'espedes et al. \cite{CNT_prox}, who
measured induced magnetic moments in excess of 0.1$\mu_\mathrm{B}$ per carbon atom in contact,
in good agreement with the theoretical prediction. The experiment consists in taking AFM and MFM
images of nanotubes on various surfaces. The difference between the topographic (AFM)
and the magnetic images (MFM) is a direct measure of the stray field coming from the nanotube
and therefore provides evidence for the induced magnetic moment (see figures \ref{fig22} and \ref{fig23}). 

In the another experiment, Mertins et al. \cite{Mertins} produced a multilayer of thin, alternating 
iron and carbon layers, of thickness 2.55 and 0.55 nm, respectively.  Then, they probed locally
the magnetic moment of the carbon by X-ray magneto-optical reflectivity of polarized synchrotron 
radiation. In this type of measurement the Fe and C absorption edges differ by about 500~eV
enabling one of establishing with precision whether the magnetic moment comes from C or not. 
With this method magnetic moments of the order of 0.05~$\mu_\mathrm{B}$ were found.

Finally we wish to mention a few experiments where ferromagnetism is claimed in heterostructures
in which none of the components is ferromagnetic. This is for instance the case of gold surfaces
and nanoparticles coated with different organic molecules \cite{Naaman,crespo,yamamoto}.
Common features to these heterostructures are the extreme magnetic anisotropy and the fact that
the magnetization is almost independent from the temperature. A possible explanation is that 
the charge transfer between the molecule and the substrate and the peculiar 2D properties of the 
organic layers result in the formation of triplet states with consequent boson 
condensation \cite{vareg}. In addition a crucial r\^ole of spin-orbit interaction has been suggested
\cite{hernie}. To our knowledge, no first principles calculations of these systems have been performed.

\section{Conclusions}

We have reviewed the most recent advances in spin-transport through organic molecules. This
is a new challenging field where disciplines such as physics, chemistry, biology and electronic engineering
are rapidly converging. In particular we have discussed the main advantages and prospectives of
spin-phenomena at the molecular level. 

From a theoretical side {\it ab initio} methods for quantum transport are rapidly approaching the limit where 
quantitative predictions of molecular transport can be made. Spin-transport however brings additional complexity 
since magnetism and strong electron correlation must be considered. These call for even more sophisticated
algorithms capable of high accuracy and of undemanding scaling with the system size, an enormous challenge for the future. Finally more and more experiments are appearing where the interaction between conducting electrons and internal molecular degrees of freedom are important. Such systems go way beyond our present computational capabilities in terms of {\it ab initio} theories and open a completely unexplored way.

\section*{Acknowledgement}
We wish to thank Cormac Toher, Ivan Rungger, Chaitanya Das Pemmaraju and 
Miguel Afonso Oliveira for useful discussions and Hang Guo for giving us access 
to unpublished material. Figures \ref{fig22} and \ref{fig23} are courtesy of J.M.D.~Coey and O.~C\'espedes.
This work is sponsored by Science Foundation of Ireland under the
grants SFI02/IN1/I175 and SFI05/RFP/PHY0062.

%

\begin{thebibliography}{99}
\bibitem{gap} Note that this is the DFT-LDA gap, which might be substantially smaller than 
the actual gap.
\end{thebibliography}

\end{document}